\documentclass{article}

\usepackage[a4paper,top=2cm,bottom=2cm,left=3cm,right=3cm,marginparwidth=1.75cm]{geometry}

\usepackage{placeins}
\usepackage{amssymb}
\usepackage{amsmath}
\usepackage{siunitx}
\PassOptionsToPackage{hyphens}{url}\usepackage{hyperref}
\usepackage{cleveref}
\usepackage[utf8]{inputenc}
\usepackage[right]{lineno}
\usepackage{csquotes}
\usepackage{booktabs}
\usepackage{longtable}
\usepackage{adjustbox}
\usepackage{array}
\usepackage{url}
\usepackage{titlesec}
\usepackage{authblk}
\usepackage{xcolor} 

\usepackage{algorithm}
\usepackage{algorithmicx}
\usepackage{algpseudocode}

\usepackage{graphicx}
\usepackage{subcaption}

\usepackage[english]{babel}
\usepackage[style=numeric, backend=biber, sorting=none]{biblatex}
\DeclareNameAlias{sortname}{family-given}
\DeclareNameAlias{default}{family-given}

\renewbibmacro{in:}{}
\DeclareFieldFormat[article]{title}{\mkbibquote{#1}\addcomma}
\DeclareFieldFormat[book]{title}{\mkbibemph{#1}\addcomma}
\DeclareFieldFormat[bookinbook]{title}{\mkbibemph{#1}\addcomma}
\DeclareFieldFormat[inbook]{title}{\mkbibquote{#1}\addcomma}
\DeclareFieldFormat[incollection]{title}{\mkbibquote{#1}\addcomma}
\DeclareFieldFormat[inproceedings]{title}{\mkbibquote{#1}\addcomma}
\DeclareFieldFormat[manual]{title}{\mkbibemph{#1}\addcomma}
\DeclareFieldFormat[misc]{title}{\mkbibemph{#1}\addcomma}
\DeclareFieldFormat[thesis]{title}{\mkbibemph{#1}\addcomma}
\DeclareFieldFormat[unpublished]{title}{\mkbibquote{#1}\addcomma}
\DeclareFieldFormat[patent]{title}{\mkbibemph{#1}\addcomma}
\DeclareFieldFormat[report]{title}{\mkbibemph{#1}\addcomma}
\DeclareFieldFormat[online]{title}{\mkbibquote{#1}\addcomma}
\DeclareFieldFormat[software]{title}{\mkbibemph{#1}\addcomma}
\DeclareFieldFormat[booklet]{title}{\mkbibemph{#1}\addcomma}
\DeclareFieldFormat[periodical]{title}{\mkbibemph{#1}\addcomma}
\DeclareFieldFormat[standard]{title}{\mkbibemph{#1}\addcomma}

\DeclareFieldFormat[article]{journaltitle}{\iffieldundef{shortjournal}{\mkbibemph{#1}\addcomma}{\mkbibemph{\printfield{shortjournal}}\addcomma}}
\DeclareFieldFormat{volume}{\bibstring{volume}~#1}
\DeclareFieldFormat{number}{\bibstring{number}~#1}

\DefineBibliographyStrings{english}{
  volume = {Vol.},
  number = {No.},
}

\renewbibmacro*{volume+number+eid}{%
  \printfield{volume}%
  \setunit*{\addspace}%
  \printfield{number}%
  \setunit{\addcomma\space}%
  \printfield{eid}}

\renewbibmacro*{journal+issuetitle}{%
  \usebibmacro{journal}%
  \setunit*{\addcomma\space}%
  \usebibmacro{volume+number+eid}%
  \setunit{\addcomma\space}%
  \usebibmacro{issue+date}}

\renewbibmacro*{publisher+location+date}{%
  \printlist{publisher}%
  \iflistundef{location}
    {\setunit*{\addcomma\space}}
    {\setunit*{\addcolon\space}}%
  \printlist{location}%
  \setunit*{\addcomma\space}%
  \usebibmacro{date}}

\DeclareCiteCommand{\cite}[\mkbibbrackets]
  {\usebibmacro{prenote}}
  {\usebibmacro{citeindex}%
   \printfield{labelnumber}}
  {\multicitedelim}
  {\usebibmacro{postnote}}

\DeclareBibliographyDriver{article}{%
  \printnames{author}%
  \setunit{\addcomma\space}%
  \printfield{title}%
  \setunit{\addcomma\space}%
  \printfield{journaltitle}%
  \setunit{\addcomma\space}%
  \printfield{volume}%
  \printfield{number}%
  \setunit{\addcomma\space}%
  \printfield{pages}%
  \setunit{\addperiod\space}%
  \printfield{year}%
  \setunit{\addspace}%
  \printfield{note}%
  \finentry
}

\DeclareFieldFormat{year}{#1}

\AtEveryBibitem{
  \clearfield{month}
  \clearfield{day}
  \ifentrytype{book}{
    \clearlist{location}
  }{}
}

\DefineBibliographyStrings{english}{
  andothers = {\textit{et al.},}
}
\DeclareFieldFormat[article]{volume}{Vol\adddot\space#1}
\DeclareFieldFormat[article]{number}{No\adddot\space#1}
\DeclareFieldFormat{pages}{pp\adddot\space#1}

\DeclareFieldFormat{url}{available at\addcolon\space\url{#1}}
\DeclareFieldFormat{urldate}{\mkbibparens{accessed \addspace#1}}

\DeclareFieldFormat{urldate}{%
  \mkbibparens{accessed\space%
    \thefield{urlday}\addspace%
    \mkbibmonth{\thefield{urlmonth}}\addspace%
    \thefield{urlyear}}}

\crefformat{figure}{#2Fig.~#1#3}
\Crefformat{figure}{#2Fig.~#1#3}
\crefformat{table}{#2Tab.~#1#3}
\Crefformat{table}{#2Tab.~#1#3}
\crefformat{section}{#2Section~#1#3}
\Crefformat{section}{#2Section~#1#3}
\usepackage{xcolor}
\newcommand{\hs}[1]{{\color{black}{#1}}}
\author[1]{Haoran Su}
\author[1]{Joseph Y.J. Chow}

\affil[1]{C2SMARTER Center, Department of Civil and Urban Engineering, New York University Tandon School of Engineering}

\title{Intersection-Aware Assessment of EMS Accessibility in NYC: A Data-Driven Approach}
\begin{document}
\maketitle

\begin{abstract}
Emergency response times are critical in densely populated urban environments like New York City (NYC), where traffic congestion significantly impedes emergency vehicle (EMV) mobility. This study introduces an intersection-aware emergency medical service (EMS) accessibility model to evaluate and improve EMV travel times across NYC. Integrating intersection density metrics, road network characteristics, and demographic data, the model identifies vulnerable regions with inadequate EMS coverage. The analysis reveals that densely interconnected areas, such as parts of Staten Island, Queens, and Manhattan, experience significant accessibility deficits due to intersection delays and sparse medical infrastructure. To address these challenges, this study explores the adoption of EMVLight, a multi-agent reinforcement learning framework, which demonstrates the potential to reduce intersection delays by 50\%, increasing EMS accessibility to 95\% of NYC residents within the critical benchmark of 4 minutes. Results indicate that advanced traffic signal control (TSC) systems can alleviate congestion-induced delays while improving equity in emergency response. The findings provide actionable insights for urban planning and policy interventions to enhance EMS accessibility and ensure timely care for underserved populations.

\end{abstract}

\section{Introduction}

\label{sec:introduction}
Emergency response times in urban areas, particularly in densely populated metropolitan regions such as New York City (NYC), are profoundly influenced by traffic congestion. Research has established that reducing the travel time of emergency vehicles (EMVs)—a critical component of overall emergency response time (ERT)—can significantly improve outcomes for urgent incidents, including emergency medical services (EMS)~\cite{blackwell2002response, SCHUSTER2024104017}, fires~\cite{jaldell2017important,challands2010relationships}, and police activities~\cite{blanes2018effect}. However, as New York City continues to undergo rapid urbanization, the growing congestion on its roadways has exacerbated this challenge. Recent reports indicate that the city's emergency response time for life-threatening situations has increased by 29\% over the past decade, reflecting a concerning trend of slower responses in high-priority cases~\cite{EMS1_NYC_Response_Times}. According to the Mayor’s Management Report, response times in New York City have risen by 82 seconds—16.1\% higher than the 8 minute and 28 second average in fiscal year 2019—surpassing even the elevated times seen during the onset of the COVID-19 pandemic in fiscal 2020, despite a 52\% increase in civilian fire deaths in fiscal 2023~\cite{NYPost_Response_Times_2023}.

Various approaches have been introduced to mitigate delays in EMV travel time, such as traffic signal preemption and route optimization. However, these methods are typically implemented in isolation and fail to address the dynamic and evolving nature of urban traffic congestion. For instance, signal preemption strategies prioritize EMVs by altering traffic signals, but often at the cost of significant disruption to non-EMV traffic. These inefficiencies propagate through the network, creating broader delays and unintended consequences for overall traffic management. Moreover, these traditional methods are static and lack the adaptability needed to respond to real-time fluctuations in traffic conditions and congestion patterns. As a result, their long-term effectiveness is limited, particularly in a complex urban environment like New York City, where traffic patterns are unpredictable and can change rapidly. Addressing these challenges necessitates more integrated and adaptive systems that holistically coordinate traffic management, minimizing disruption to non-EMV traffic while ensuring swift emergency response. 

A recent study on emergency response accessibility in NYC identified significant disparities in service coverage, particularly in underserved areas such as Staten Island and parts of Queens~\cite{chung2024access}. These regions, referred to as "emergency service deserts," experience prolonged response times due to inadequate proximity to fire stations and EMS. The study further highlights a correlation between population density and emergency service accessibility, with less dense areas facing longer delays. While these findings emphasize the need for innovative traffic management solutions to enhance emergency response across all boroughs, the proposed accessibility model in~\cite{chung2024access} has been shown to underestimate EMV travel times within the city. An improved EMS accessibility model is therefore essential to accurately identify and address vulnerable areas in NYC.

Building upon these findings, this study aims to develop an EMS accessibility model that incorporates the potential adoption of large-scale traffic signal control (TSC) systems to facilitate EMV passage in congested networks. Using this EMS accessibility model, we can simulate travel times from the nearest medical facility to any location in the city, enabling the identification of vulnerable regions in terms of EMS coverage. Additionally, this model allows for the evaluation of how a TSC scheme, such as \textit{EMVLight}~\cite{su2023emvlight}, could support these vulnerable areas. By integrating advanced modeling capabilities with insights from the accessibility study, this study is able to demonstrate the potential to significantly enhance emergency response times in NYC’s most vulnerable and congested regions.

This study makes the following contributions:
\begin{enumerate}
    \item It integrates the New York City road network, intersection data, and population distribution into a unified analytical framework.
    \item It proposes an EMS accessibility model that accounts for the impact of intersection density along emergency routes.
    \item It identifies vulnerable regions using the proposed model, investigates the EMS accessibility and illustrates the potential improvements in accessibility achievable through the incorporation of \textit{EMVLight}.
\end{enumerate}

The structure of this study is as follows. Section~\ref{sec:literature_review} reviews existing literature on traffic signal controls for EMVs, Intersection delay and EMS accessibility. The datasets employed in this study are introduced in Section~\ref{sec:methodology}, which also presents the proposed intersection-aware EMS accessibility model. Section~\ref{sec:results} details the travel time analysis and the vulnerable regions based on the proposed method. Demographic evaluations are conducted on the result in Section~\ref{sec:discussion} and an adoption of \textit{EMVLight} is carried out in Subsec.~\ref{subsec:improvement}, highlighting the areas of New York City that stand to benefit the most. Finally, conclusions are summarized in Section~\ref{sec:conclusion}.

\section{Related Works}
\label{sec:literature_review}
This section reviews key literature supporting our study. Subsection~\ref{tsc_for_emvs} examines advancements in traffic signal control (TSC) for EMVs, Subsection~\ref{subsec:intersection_delay} focuses on intersection delays as a critical factor in EMV travel times and Subsection~\ref{subsec:ems_accessibility} explores EMS accessibility and its disparities. Together, these insights provide the foundation for our proposed methodology.

\subsection{EMS accessibility}\label{subsec:ems_accessibility}
The accessibility of EMS is a critical component of public health and safety, particularly in urban areas where rapid response times are essential for saving lives. Studies have developed a variety of methodologies to evaluate and improve EMS accessibility. Novak and Sullivan~\cite{novak2014link} proposed a link-focused approach for assessing access to emergency services, emphasizing the importance of connectivity within urban road networks. Similarly, Xia et al.~\cite{xia2019measuring} utilized big GPS data to measure spatiotemporal accessibility, highlighting the dynamic nature of EMS access influenced by traffic and temporal factors.

Geographic Information Systems (GIS) have played a pivotal role in analyzing and addressing spatial disparities in EMS accessibility. Tansley et al.~\cite{tansley2015spatial} employed GIS to assess spatial access to EMS in low- and middle-income countries, revealing significant geographic inequities. Hashtarkhani et al.~\cite{hashtarkhani2020age} introduced an age-integrated GIS-based methodology to evaluate EMS accessibility, demonstrating the potential to tailor EMS services to different demographic groups. In Dhaka, Bangladesh, Ahmed et al.~\cite{ahmed2019impact} examined the impact of traffic variability on EMS access, highlighting the challenges posed by urban congestion in providing equitable emergency healthcare.

Several studies have addressed the influence of environmental and infrastructural factors on EMS accessibility. Green et al.~\cite{green2017city} explored city-scale accessibility during flood events, underscoring the importance of resilient infrastructure to maintain EMS operations during natural disasters. Similarly, Yin et al.~\cite{yin2017evaluating} evaluated the cascading impacts of sea-level rise and coastal flooding on EMS access in Lower Manhattan, New York City, revealing vulnerabilities in emergency response under climate-induced hazards. Albano et al.~\cite{albano2014gis} developed a GIS-based model to estimate the operability of emergency response structures during floods, providing critical insights for urban planning in disaster-prone areas.

Addressing urban-rural disparities has also been a key focus in EMS accessibility research. Luo et al.~\cite{luo2022locating} examined strategies for locating EMS facilities to reduce urban-rural inequalities, while Carr et al.~\cite{carr2009access} highlighted access challenges in rural areas within the United States. Holguin et al.~\cite{holguin2018access} proposed urban planning methodologies aimed at generating equity in EMS access, emphasizing the need for targeted interventions in underserved regions. The spatiotemporal aspect of EMS accessibility has been further explored in studies like those by Hu et al.~\cite{hu2020impact}, who analyzed the impact of traffic on spatial accessibility variations in inner-city Shanghai. Shi et al.~\cite{shi2022spatial} conducted a spatial accessibility assessment of urban tourist attractions in Shanghai, demonstrating the need for comprehensive urban planning to address emergency response challenges in high-density areas.

Chung et al.~\cite{chung2024access} present a geospatial analysis of emergency service accessibility in New York City, identifying significant disparities in coverage, particularly in Staten Island and parts of Queens. The study introduces an accessibility index that combines spatial proximity, road network connectivity, and response time data to highlight regions with low emergency service availability. Additionally, the authors examine the impact of traffic congestion on EMS response times, emphasizing challenges in densely populated areas.

In summary, EMS accessibility is influenced by a combination of spatial, temporal, and demographic factors. The integration of advanced GIS techniques, real-time traffic data, and tailored urban planning approaches is critical for addressing disparities and ensuring equitable access to emergency services across diverse geographic and socio-economic contexts.
\subsection{Intersections delay}\label{subsec:intersection_delay}
The influence of intersection density and topology on travel time has been a subject of extensive research. Higher intersection densities often lead to increased delays due to frequent stops and interruptions in traffic flow. Ewing and Cervero's meta-analysis~\cite{EwingCervero2010} established that although higher intersection densities are associated with reduced vehicle miles traveled (VMT), they often result in longer travel times due to frequent interruptions. Al-Dabbagh et al.~\cite{al2019impact} further emphasize the role of intersection topology, demonstrating how intersection design impacts traffic congestion in urban cities. The \textit{Highway Capacity Manual} (HCM)~\cite{HCM2010} offers methodologies for assessing urban street performance, highlighting the influence of signalized intersections on travel time. 

Specific studies have also examined how intersections affect travel times across different transportation modes. Feng et al.~\cite{feng2015quantifying} quantified the joint impact of stop locations, signalized intersections, and traffic conditions on bus travel times, highlighting the compounded delays experienced by public transport. Strauss and Miranda-Moreno~\cite{strauss2017speed} explored cyclist travel patterns, demonstrating how delays at intersections significantly influence travel times in the Montreal network. Furthermore, Park et al.~\cite{park2016measuring} utilized Bluetooth-based data to measure intersection performance, illustrating how advanced data collection methods can provide detailed insights into intersection-related delays. Ivanov et al.~\cite{AssessingTrafficCapacity2022} also highlight the importance of intersection geometry and control mechanisms in mitigating delays in high-density areas. These findings collectively underscore the critical role of intersection design and performance in influencing travel time and highlight the need for targeted urban planning interventions. 

When it comes to EMV travel time, reducing delays at intersections is particularly critical, as every second lost can directly impact emergency outcomes. Therefore, the ability to dynamically manage and optimize intersection performance is essential for ensuring timely emergency response, forming the focus of this study.

\subsection{TSC for EMVs}\label{tsc_for_emvs}
The optimization of TSC for EMS is paramount in reducing response times and enhancing public safety. Traditional TSC methods, such as signal preemption, temporarily adjust traffic signals to grant EMVs priority passage through intersections~\cite{obeck1991traffic}. While effective in facilitating EMV movement, these approaches often disrupt the flow of non-EMV traffic, leading to broader network inefficiencies~\cite{hashim2013traffic}. Advancements in intelligent transportation systems (ITS) have introduced more adaptive and dynamic TSC strategies. Bhate et al.~\cite{bhate2018iot} proposed an IoT-based intelligent traffic signal system that leverages real-time data to prioritize EMVs, demonstrating significant reductions in response times. Similarly, Noori et al.~\cite{noori2016connected} developed a connected vehicle-based TSC strategy, utilizing vehicle-to-infrastructure communication to enhance EMV preemption efficiency.
Predictive control models have also been explored to anticipate EMV arrivals and adjust signal timings accordingly. Qin and Khan~\cite{qin2012control} investigated control strategies for traffic signal timing transitions during EMV preemption, highlighting methods to minimize disruptions to regular traffic flow. Additionally, Shanmughasundaram et al.~\cite{shanmughasundaram2018li} introduced a Li-Fi-based automatic traffic signal control system, offering a novel approach to EMV prioritization through high-speed data transmission.

The integration of multi-agent reinforcement learning (MARL) frameworks has further advanced TSC for EMVs. Su et al.~\cite{su2023emvlight} presented EMVLight, a decentralized MARL framework that optimizes both EMV routing and traffic signal coordination, effectively reducing response times while maintaining overall traffic efficiency. This approach underscores the potential of machine learning techniques in addressing the complexities of urban traffic management.

Comprehensive reviews by Bin Wan Hussin et al.~\cite{bin2019review} and Yu et al.~\cite{yu2022state} provide extensive analyses of existing TSC techniques for EMVs, encompassing traditional methods and emerging technologies. These reviews offer valuable insights into the evolution of TSC strategies and underscore the necessity for continued innovation in this domain.
\section{Methodology}
\label{sec:methodology}
In this section, we introduce the datasets used for this study in Subsec.~\ref{subsec:datasets} and the EMS accessibility model in \hs{Subsec.~\ref{subsec:ems_model}.}
\subsection{Datasets}\label{subsec:datasets}
In this subsection, we introduce the datasets used for the proposed methodology.
\subsubsection{Emergency medical services sites}
\begin{figure}[h!]
    \centering
    \includegraphics[width=\textwidth]{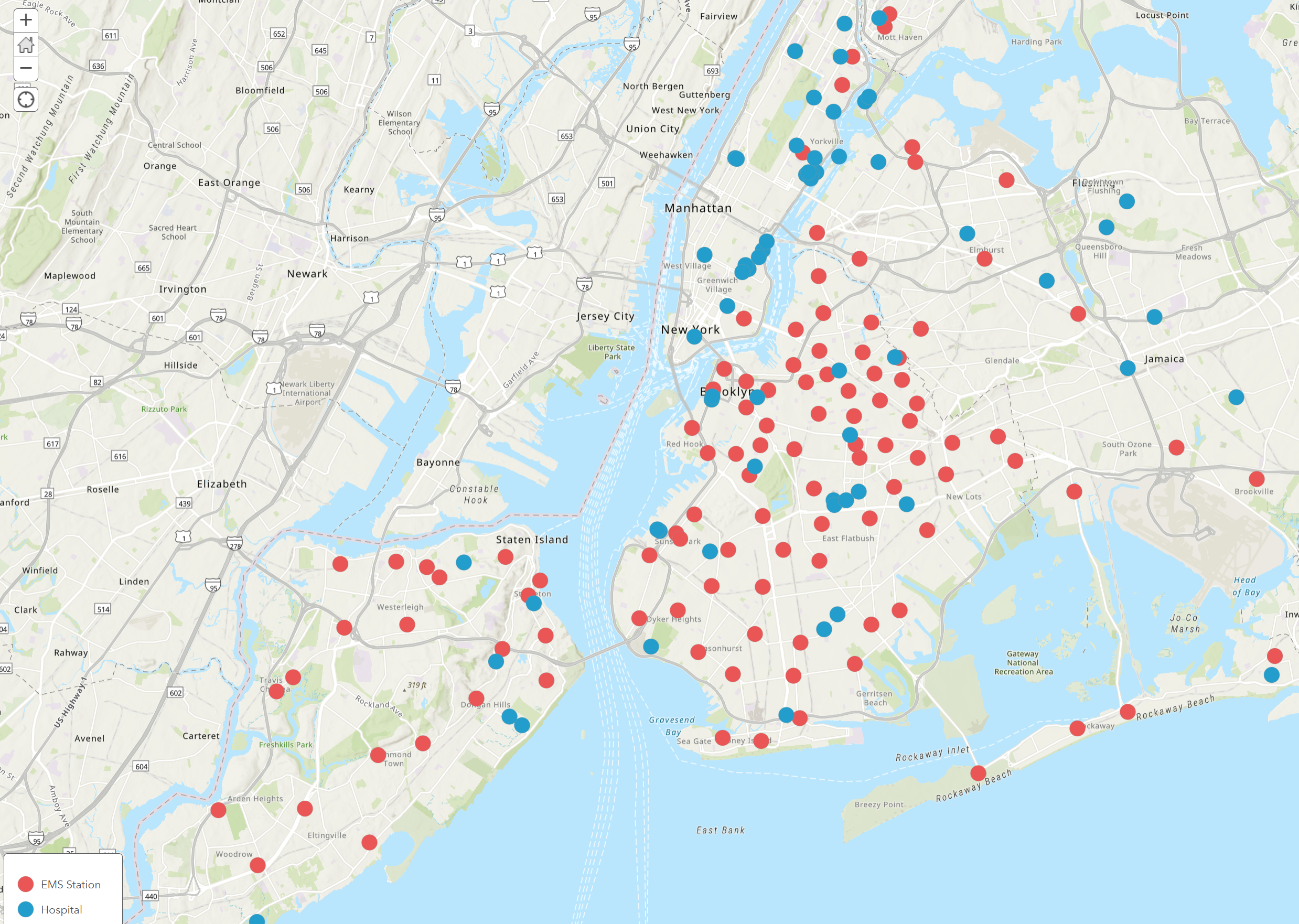} 
    \caption{EMS Stations (Blue) and Hospitals (Red) in NYC.}
    \label{fig:hospitals_and_ems_stations}
\end{figure}

The locations of emergency service sites are sourced from the Homeland Infrastructure Foundation-Level Data (HIFLD) Open Data repository. Given that this study concentrates on optimizing the transit of ambulances and other Emergency Medical Services (EMS) vehicles, we specifically extract the precise locations of hospitals~\cite{HIFLD_hospital} and EMS stations~\cite{HIFLD_EMS_Stations} to ensure targeted analysis and effective modeling of emergency response routes. Note that the EMS stations dataset also includes certain fire stations, as these facilities provide EMS services as well. The numbers of sites by each borough is provided in Tab.~\ref{table:hospitals_and_ems_stations} and their locations are represented in Fig.~\ref{fig:hospitals_and_ems_stations}.
\begin{table}[ht]
\centering
\begin{tabular}{@{}lccc@{}}
\toprule
\textbf{Borough} & \textbf{EMS Stations} & \textbf{Hospitals} & \textbf{Total} \\ \midrule
Bronx            & 19                    & 12                 & 31             \\
Brooklyn         & 76                    & 15                 & 91             \\
Manhattan        & 28                    & 23                 & 51             \\
Queens           & 15                    & 10                 & 25             \\
Staten Island    & 23                    & 6                 & 29             \\ \midrule
\textbf{Total NYC} & \textbf{161}        & \textbf{66}       & \textbf{227}   \\ \bottomrule
\end{tabular}
\caption{Number of EMS Stations and Hospitals by boroughs in NYC}
\label{table:hospitals_and_ems_stations}
\end{table}

\subsubsection{Road network}
\begin{figure}[h!]
    \centering
    \includegraphics[width=\textwidth]{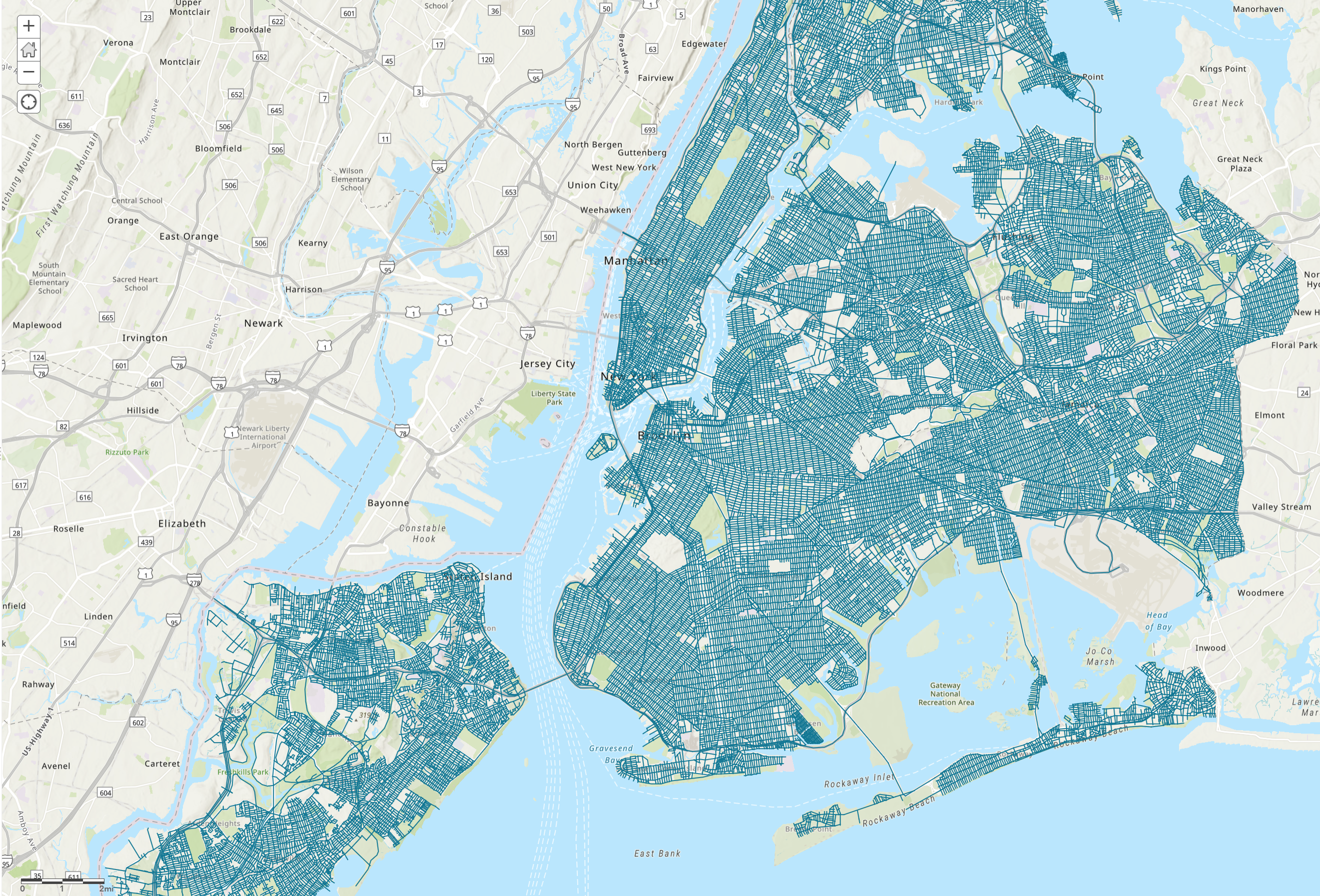} 
    \caption{All NYC streets following NYC Street Centerline.}
    \label{fig:nyc_street_centerline}
\end{figure}

OpenStreetMap (OSM) is employed to visualize the streets and intersections of New York City. The road network is represented as a symmetric multigraph, capturing the intricate connections and pathways characteristic of the urban transportation infrastructure. To ensure precision and clarity in the representation, the NYC Street Centerline (CSCL) dataset~\cite{nyc_cscl} is utilized.  An overview of the network is illustrated in Fig.~\ref{fig:nyc_street_centerline}. This representation serves as the foundation for analyzing the accessibility of EMS site locations, with travel times along the network serving as a key metric for reachability and service efficiency. By employing this approach, the study offers a realistic depiction of urban mobility and its implications for emergency response performance.

\subsubsection{Signalized intersections}
\begin{figure}[h!]
    \centering
    \includegraphics[width=\textwidth]{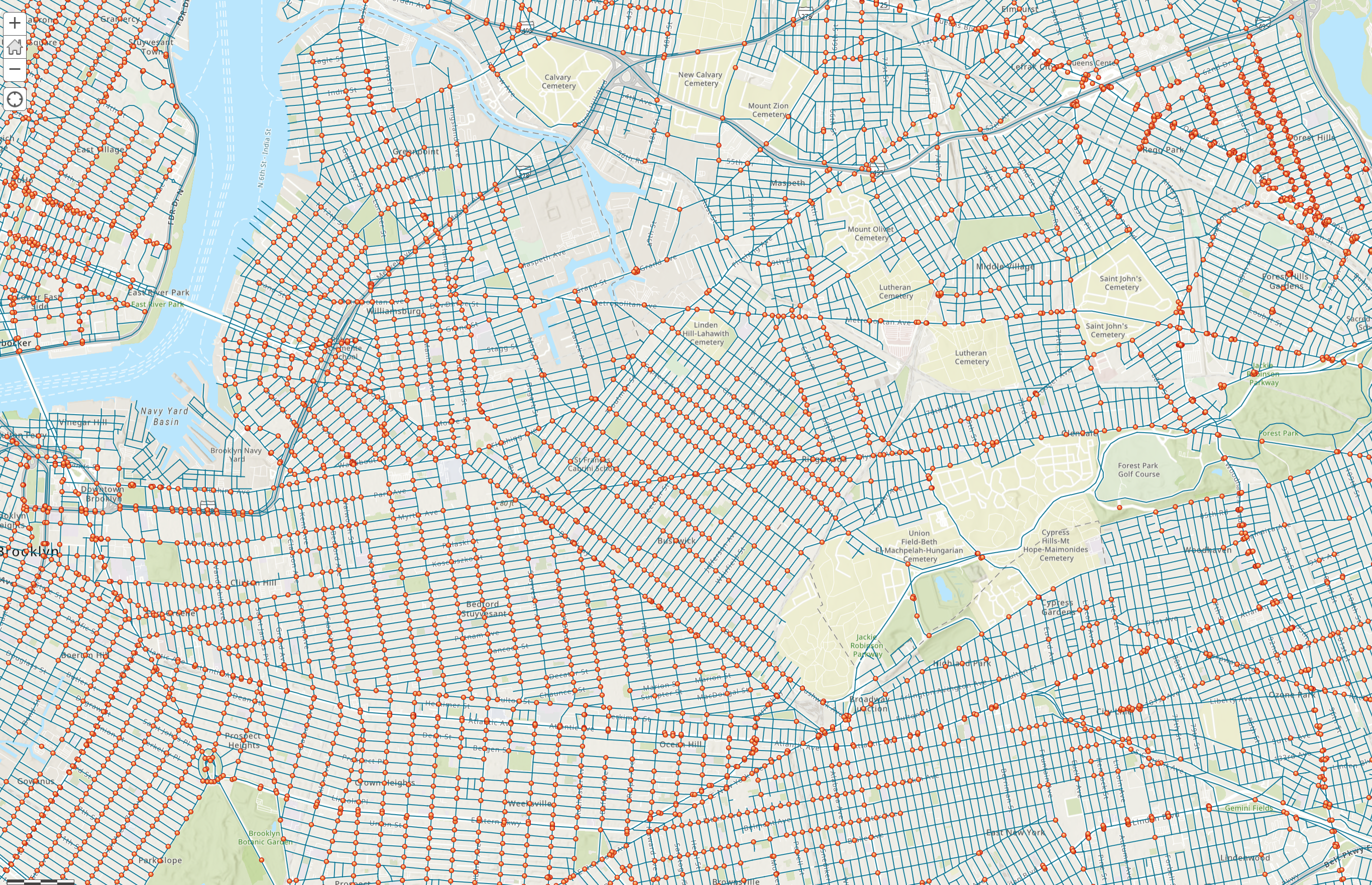}
    \caption{Signalized intersections layout in Brooklyn and Queens}
\label{fig:signalized_intersections}
\end{figure}

Currently, there is no publicly available open-source dataset that comprehensively identifies all signalized intersections in New York City. However, we have transformed the NYC Street Centerline (CSCL) shapefile to generate an approximate map of signalized intersection locations. Following pre-processing of the CSCL data, we filtered the dataset to include only road types classified as streets or highways and restricted traffic directions to multi-way configurations. Intersection points were derived by extracting line endpoints from the filtered road geometries. To address precision errors and reduce overlap caused by minor positional discrepancies, the coordinates of these endpoints were rounded. Duplicate intersection points were subsequently removed, resulting in a refined and clean dataset of potential signalized intersection locations. Refer to Fig.~\ref{fig:signalized_intersections} for an illustrative visualization of the signalized intersection map in New York City, where blue lines represent streets or highways as part of the network in Subsec.~\ref{subsubsec:graph} and orange dots represent traffic lights.

Given that each signalized intersection can function as an agent within a large-scale TSC system such as \textit{EMVLight}, we regard this map as a realistic near-future representation of the network for facilitating EMV passage in NYC.\footnote{\hs{This intersections map is available at: \url{https://zenodo.org/records/14280428}}}

\subsubsection{NYC census data}
\begin{figure}[h!]
    \centering
    \includegraphics[width=\textwidth]{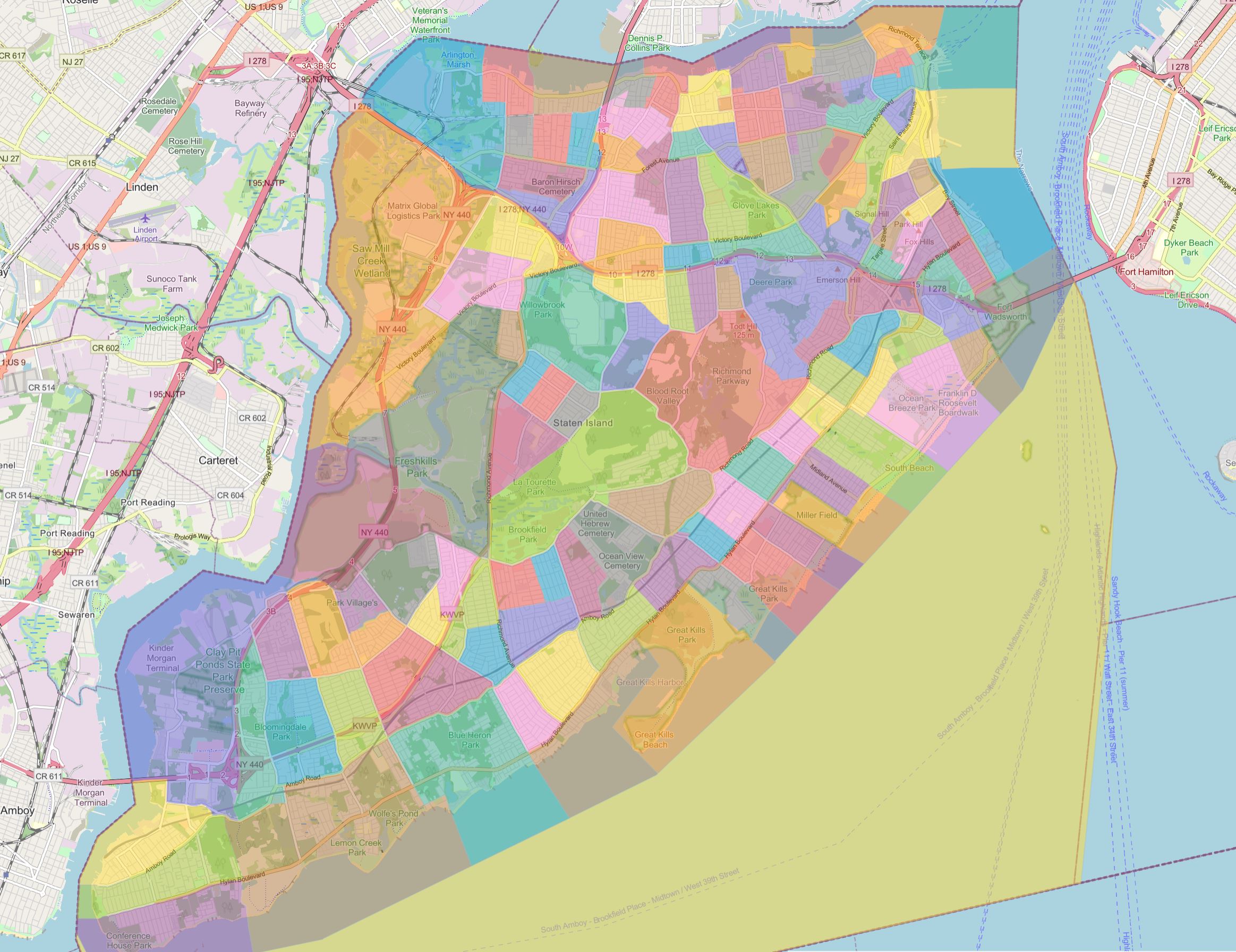} 
    \caption{Census tracts in Staten Island. Each polygon represents a census tract.}
    \label{fig:staten_island_census_tracts}
\end{figure}

\begin{figure}[h!]
    \centering
    \includegraphics[width=\textwidth]{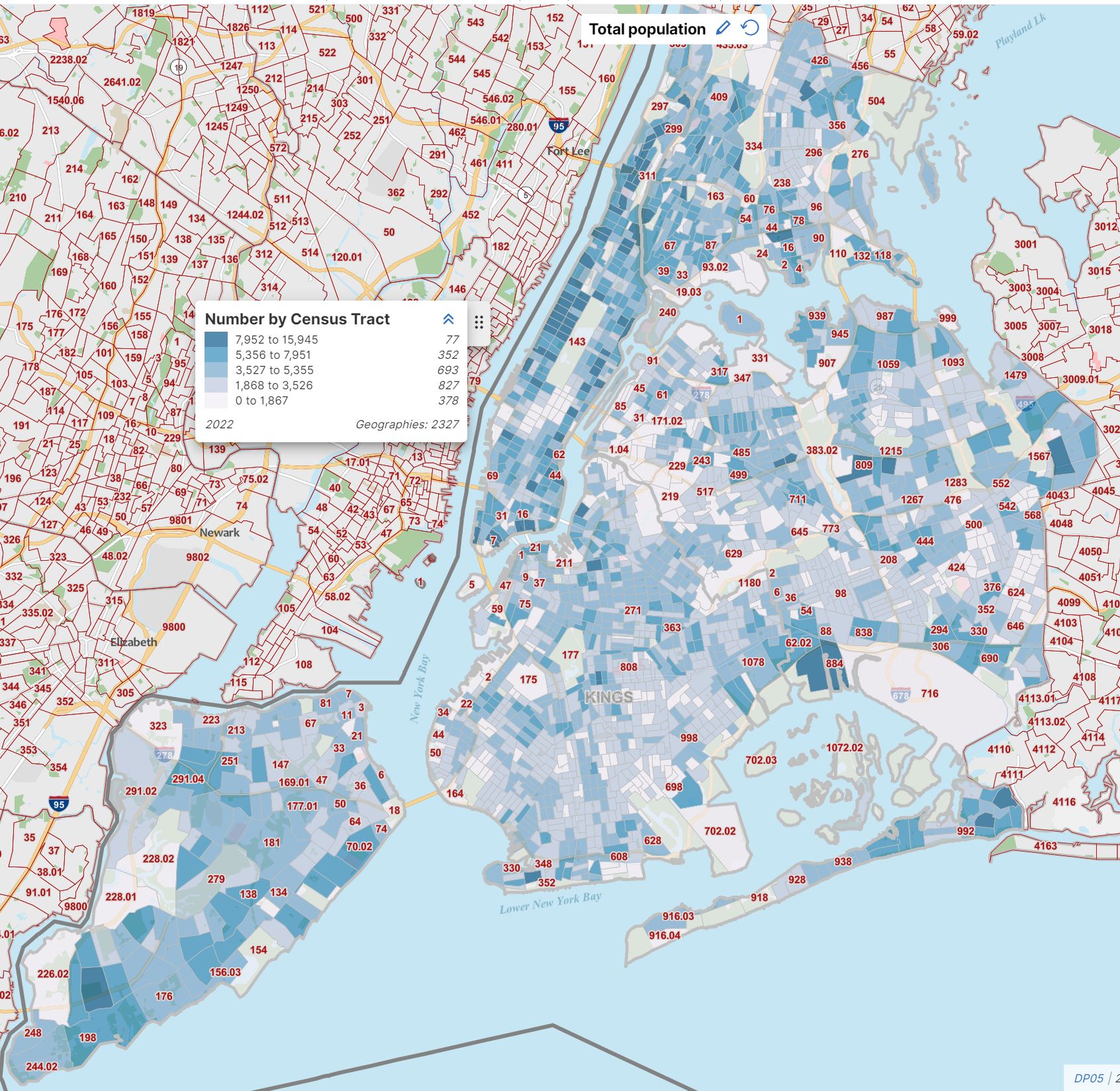} 
    \caption{Population distribution of NYC census tracts.}
    \label{fig:total_population}
\end{figure}

Census data plays a pivotal role in conducting analyses of accessibility and vulnerability. While \hs{Chung ~\cite{chung2024access}} employs a five-year average for such assessments, this study utilizes 2023 population data at the census tract level provided by the U.S. Census Bureau~\cite{bureau2023census}, thereby enhancing temporal precision. The spatial configurations of census tracts are sourced from the U.S. Census Bureau's dataset~\cite{us_census_bureau_census_tracts_2023}, with a representative example for Staten Island illustrated in Fig.~\ref{fig:staten_island_census_tracts}. 
The demographic data are DP05 from American Community Survey (ACS) Demographic and Housing Estimate~\cite{acs2021demographic} 2022 5-year estimate data profiles. In subsequent sections of this study, we will undertake a detailed examination of the spatial distribution of various demographic characteristics across New York City. This analysis aims to provide deeper insights into issues surrounding EMS accessibility. To provide an overview of the census data, a population density map illustrating the spatial distribution of the population is presented in Fig.~\ref{fig:total_population}.

\subsection{Intersection-aware EMS Accessibility Model}\label{subsec:ems_model}
\hs{Accessibility is commonly defined as the ease of reaching desired destinations, with measures classified into infrastructure-based, location-based, person-based, and utility-based categories~\cite{geurs2004accessibility}. Among these, location-based measures, particularly travel time within a predefined threshold, are widely used for their simplicity and relevance to EMS planning. Travel time thresholds, such as the NFPA 1710 standard of \(\tau = 4\) minutes~\cite{NFPA1710}, serve as benchmarks for evaluating whether populations can access critical EMS facilities.

This study adopts a travel time-based measure of accessibility, incorporating intersection density to account for delays in densely connected urban areas. Building on the methodology proposed by Chung et al.~\cite{chung2024access}, the following subsections outline the model’s graph representation, travel time assignment, and population weighting framework, providing a refined evaluation of EMS accessibility in NYC.}
\subsubsection{Graph representation}\label{subsubsec:graph}
The road network is represented as a graph \( G = (V, E) \), where \( V \) is the set of nodes, each corresponding to either a road intersection or endpoint, and \( E \) denotes the set of edges, each representing a distinct road segment connecting two nodes. Attributes associated with each edge, such as road length and speed limit, facilitate the initial computation of travel time across the network under idealized conditions.
\subsubsection{Travel time assignment}
To capture the influence of urban intersection density on travel times, an intersection density metric \( I(v) \) is defined for each node \( v \in V \). This metric is calculated as the number of intersections within a fixed radius \( r \) around the node, normalized by the area of the radius, such that:
\begin{equation}
    I(v) = \frac{\text{Number of intersections within radius } r}{\pi r^2}.
\end{equation}
A higher \( I(v) \) value implies a denser concentration of intersections, which is expected to contribute to travel delays, particularly affecting emergency response times in congested urban zones. Based on Leadership in Energy and Environmental Design (LEED)~\cite{LEED_IntersectionDensity}, we adopt \( r = 800 \, \mathrm{m} \) for intersection density evaluations.

To compute travel times in the road network, we follow the graph representation introduced in Subsec.~\ref{subsubsec:graph}. Each edge \( e \in E \) is assigned a baseline travel time \( T(e) \), calculated as the edge, i.e., road segment, length divided by the speed limit. The default speed limit on most New York City streets is \( 25 \, \mathrm{mph} \), implemented under the Vision Zero initiative~\cite{nyc_speed_limit} to enhance traffic safety. On highways and parkways, the limit is generally higher, ranging from \( 50 \, \mathrm{mph} \) to \( 55 \, \mathrm{mph} \) depending on signage. The baseline travel time matrix \( T \) is then defined as:
\begin{equation}
    T_{ij} = \text{Shortest path travel time from node } v_i \text{ to node } v_j,
\end{equation}
where \( T_{ij} \) represents the cumulative travel time across all edges along the shortest path between nodes \( v_i \) and \( v_j \), under traffic-free conditions (no delays). 

To account for delays caused by intersection density, we adjust the travel time for each edge \( e \) using a node-based intersection delay design. The delay for an edge \( e \) connecting nodes \( v_m \) and \( v_n \) is modeled as:
\begin{equation}
    D_\text{intersection}(e) = \alpha \cdot \frac{I(v_m) + I(v_n)}{2},
\end{equation}
where \( I(v_m) \) and \( I(v_n) \) are the intersection densities of the nodes at the two ends of the edge \( e \), and \( \alpha \) is a calibration factor reflecting the impact of intersection density on delays, in units of [minutes $\cdot$ square meters per intersection]. This design ensures that travel times accurately capture the influence of densely connected nodes at the start and end of each road segment, aligning with the principle that congestion delays are more pronounced in highly interconnected regions.

The adjusted travel time for each edge is then given by:
\begin{equation}
    T'(e) = T(e) + D_\text{intersection}(e),
\end{equation}
and the corresponding adjusted travel time matrix \( T' \) is:
\begin{equation}
    T'_{ij} = \text{Shortest path travel time from node } v_i \text{ to node } v_j \text{ incorporating intersection delays.}
\end{equation}

A node \( v_i \) is considered accessible to/from an EMS site (EMS station or hospital) \( s_j \) if:
\begin{equation}
    T'_{ij} \leq \tau,
\end{equation}
where \( \tau \) is a benchmark travel time, typically set as \( 4 \, \mathrm{minutes} \), aligned with the National Fire Protection Association standards~\cite{NFPA1710}.

By integrating both road characteristics and node-based intersection delays, this framework offers a more realistic evaluation of travel times, enabling the identification of vulnerable regions where response times exceed the benchmark \( \tau \). While the model operates under the assumption that the EMV can travel unimpeded between nodes—an idealization that does not fully align with real-world conditions—it effectively captures the concept of accessibility and accounts for intersection delays along the route, providing valuable insights regarding EMS accessibility.

\subsubsection{Population assignment}\label{subsubsec:pop_assignment}
To accurately gauge the number of residents affected by EMS accessibility, population assignment is also incorporated extending the idea proposed by \cite{chung2024access}. Each node \( v \in V \) is assigned a population density by intersecting census tracts with node-based Voronoi polygons\cite{burrough20158}. Census tracts provide demographic data, including the population count \( P(c) \) and livable area \( L(c) \). 

For simplicity, the livable area of each census tract in our study is considered synonymous with its total area, which can be acquired directly from \cite{bureau2023census}.
The effective population density \( \rho^*(c) \) is defined as:
\begin{equation}
    \rho^*(c) = \frac{P(c)}{L(c)}.
\end{equation}
See Table~\ref{tab:population_density_by_borough} for total area, population and population density by each borough.
\begin{table}[ht]
    \centering
    \renewcommand{\arraystretch}{1.2} 
    \setlength{\tabcolsep}{10pt} 
    \begin{tabular}{lccc}
        \toprule
        \textbf{Borough} & \textbf{$L(c)$ [mi$^2$]} & \textbf{$P(c)$ [M]} & \textbf{$\rho^*(c)$ [k ppl/mi$^2$]} \\
        \midrule
        Bronx          & 42.2  & 1.42 & 33.65 \\
        Brooklyn       & 69.4  & 2.57 & 37.04 \\
        Manhattan      & 22.7  & 1.63 & 71.81 \\
        Queens         & 108.7 & 2.27 & 20.88 \\
        Staten Island  & 57.5  & 0.47 & 8.17  \\
        \midrule
        \textbf{Total NYC} & 300.5 & 8.36 & 27.82 \\
        \bottomrule
    \end{tabular}
    \caption{Total area, population and population density for boroughs and NYC.}
    \label{tab:population_density_by_borough}
\end{table}

The Voronoi region \( R(v) \) associated with node \( v \) defines the area influenced by that node within the road network. The population assigned to each node \( v \), \( P(v) \), is then computed by proportionally weighting the overlap of \( R(v) \) with each census tract \( c \in C \), as follows:
\begin{equation}
    P(v) = \sum_{c \in C} \left( |R(v) \cap c| \cdot \rho^*(c) \right),
\end{equation}
where \( |R(v) \cap c| \) represents the area of overlap between the Voronoi region \( R(v) \) and census tract \( c \). This assignment ensures that each node in the network has a population weighting, allowing for the assessment of accessibility impacts across densely and sparsely populated areas.

By integrating both intersection-adjusted travel times and population density metrics, the modified EMS model provides a more granular perspective on emergency accessibility in urban environments, particularly highlighting regions where dense intersections may exacerbate response delays.

\subsubsection{Vulnerability determination}
Vulnerability to insufficient EMS coverage is assessed by comparing the shortest adjusted travel time \( T'_{i,\text{EMS}} \) from each node \( v_i \) to the nearest EMS facility against a benchmark threshold \( \tau \), following National Fire Protection Association (NFPA) guidelines\cite{NFPA1710}. Nodes with \( T'_{i,\text{EMS}} > \tau \) are deemed inaccessible.

Regions are classified as “vulnerable” if they contain clusters of inaccessible nodes. To quantify the impact, population weights are assigned to each inaccessible node, allowing for an aggregate measure of residents underserved by EMS. This population-weighted vulnerability assessment highlights priority areas for potential TSC enhancement.

\section{Results}\label{sec:results}
In this section, we present travel times using the proposed EMS accessibility model outlined in Subsec.\ref{subsec:travel_time}. We then compare the simulated EMV travel time with the ground-truth travel time in Subsec.~\ref{subsec:comparison_w_real_world}. Vulnerable regions with respect to EMS accessibility are identified in Subsec.~\ref{subsec:vulnerable_regions}. 
\subsection{Travel Time}\label{subsec:travel_time}

Table~\ref{tab:travel_time_summary} summarizes the travel time statistics for EMS stations, hospitals, and a combined category considering EMS stations and hospitals as one type of facility providing comprehensive medical services based on the simulation of the proposed EMS accessibility model. The statistics are presented across different \(\alpha\) values, where \(\alpha\) is the intersection delay factor introduced in Subsec.~\ref{subsec:ems_model}, which represents the additional delay induced by intersection density. \(\alpha\) is expressed in unit of seconds $\cdot$ square meters for better readability.

Travel times are reported in minutes and include key percentiles (25\%, 50\%, 75\%, 97.5\%, and 100\%) alongside the average travel time for each category. The 97.5\% percentile represents the practically longest travel time while excluding a small number of extreme outliers in the complete simulated travel time distribution. 

Travel time to EMS stations are typically shorter due to the higher density and wider distribution of EMS stations compared to hospitals. Hospitals, although fewer in number, extend the range of travel times because of their greater average distance in urban areas. Consequently, the overall travel time to the nearest emergency medical facility is slightly shorter than that to the nearest EMS station, as the inclusion of hospitals increases the number of accessible options.
\begin{table}[ht]
    \centering
    \renewcommand{\arraystretch}{1.2} 
    \setlength{\tabcolsep}{5pt} 
    \begin{tabular}{lccccccc}
        \toprule
        \textbf{Category} & \(\boldsymbol{\alpha (\text{s}\cdot\text{m}^2)}\) & \textbf{25\%} & \textbf{50\%} & \textbf{75\%} & \textbf{97.5\%} & \textbf{100\%} & \textbf{Average} \\
        \midrule
        EMS Stations & 0   & 2.15 & 3.10 & 4.18 & 6.60 & 7.95 & 4.80 \\
                     & 5   & 2.28 & 3.23 & 4.35 & 6.85 & 8.12 & 4.97 \\
                     & 10  & 2.50 & 3.48 & 4.72 & 7.28 & 8.42 & 5.19 \\
                     & 15  & 2.95 & 3.98 & 5.12 & 7.98 & 8.93 & 5.68 \\
        \midrule
        Hospital     & 0   & 3.45 & 5.03 & 6.82 & 9.35 & 10.90 & 6.70 \\
                     & 5   & 3.68 & 5.11 & 6.93 & 9.46 & 11.15 & 6.95 \\
                     & 10  & 3.85 & 5.29 & 7.12 & 9.71 & 11.41 & 7.22 \\
                     & 15  & 4.10 & 5.81 & 7.60 & 10.32 & 11.92 & 7.50 \\
        \midrule
        Overall      & 0   & 2.07 & 2.98 & 3.92 & 6.03 & 7.22 & 3.92 \\
             & 5   & 2.18 & 3.13 & 4.08 & 6.12 & 7.34 & 4.09 \\
             & 10  & 2.35 & 3.42 & 4.47 & 6.48 & 7.52 & 4.19 \\
             & 15  & 2.73 & 3.86 & 4.99 & 7.28 & 8.38 & 4.41 \\
        \bottomrule
    \end{tabular}
    \caption{Summary of travel time [mins] for EMS stations, hospitals, and combined (Overall) given different \(\alpha\) values.}
    \label{tab:travel_time_summary}
\end{table}

\begin{figure}[H]
    \centering
    \includegraphics[width=\textwidth]{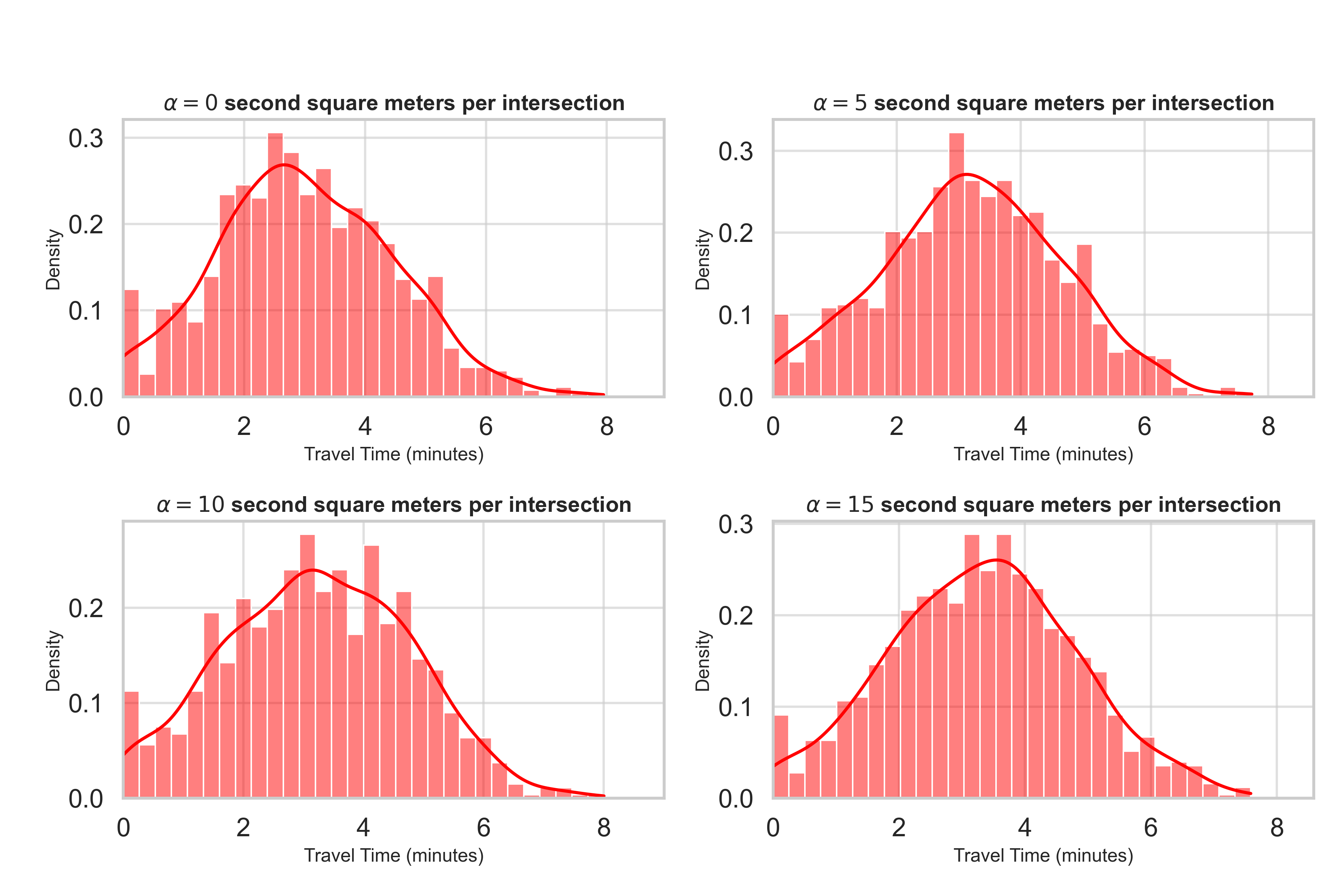}
    \caption{Travel time distributions for EMS stations across different \(\alpha\) values.}
    \label{fig:ems_travel_time}
\end{figure}

\begin{figure}[H]
    \centering
    \includegraphics[width=\textwidth]{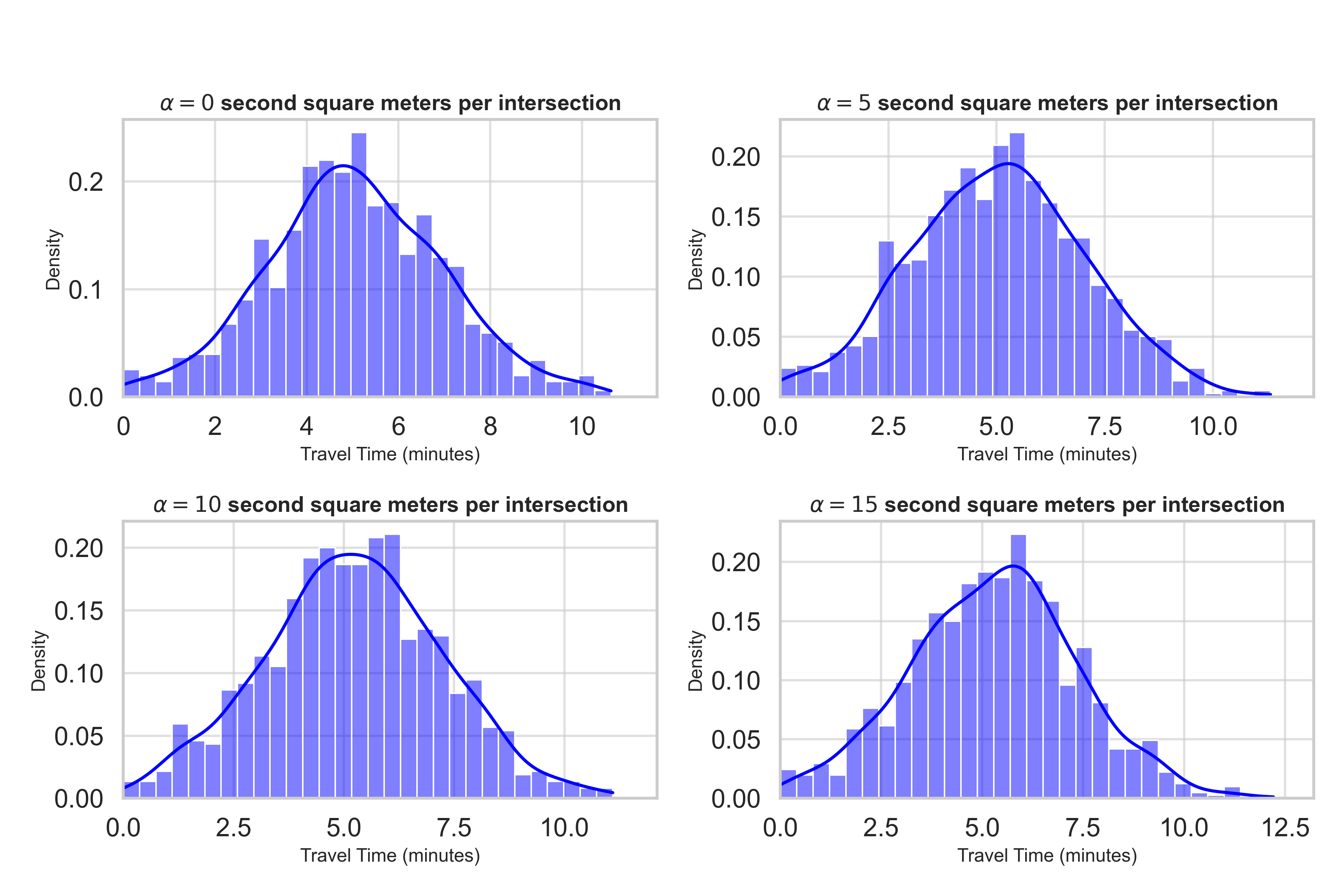}
    \caption{Travel time distributions for Hospital across different \(\alpha\) values.}
    \label{fig:hospital_travel_time}
\end{figure}

Figures~\ref{fig:ems_travel_time} and~\ref{fig:hospital_travel_time} illustrate the travel time distributions for EMS stations and hospital, respectively, under different \(\alpha\) values. Each figure consists of subplots representing travel time distributions for \(\alpha = 0, 5, 10,\) and \(15\) $\text{s}\cdot\text{m}^2$. The distributions show an increasing pattern in travel times as \(\alpha\) grows, reflecting the impact of intersection-induced delays. 

\subsection{Calibration of \texorpdfstring{\boldmath$\alpha$}{alpha}}\label{subsec:comparison_w_real_world}
To validate the correctness of the proposed model’s predicted travel time, which accounts for delays due to intersections, we summarize the EMV travel times recorded in the 911 NYC End-to-End dataset~\cite{NYC911Data}. These travel times are derived from the time difference between agency dispatch and agency arrival and span the period from the week of November 18, 2023, to October 21, 2024.
\begin{figure}[h!]
    \centering
    \includegraphics[width=\textwidth]{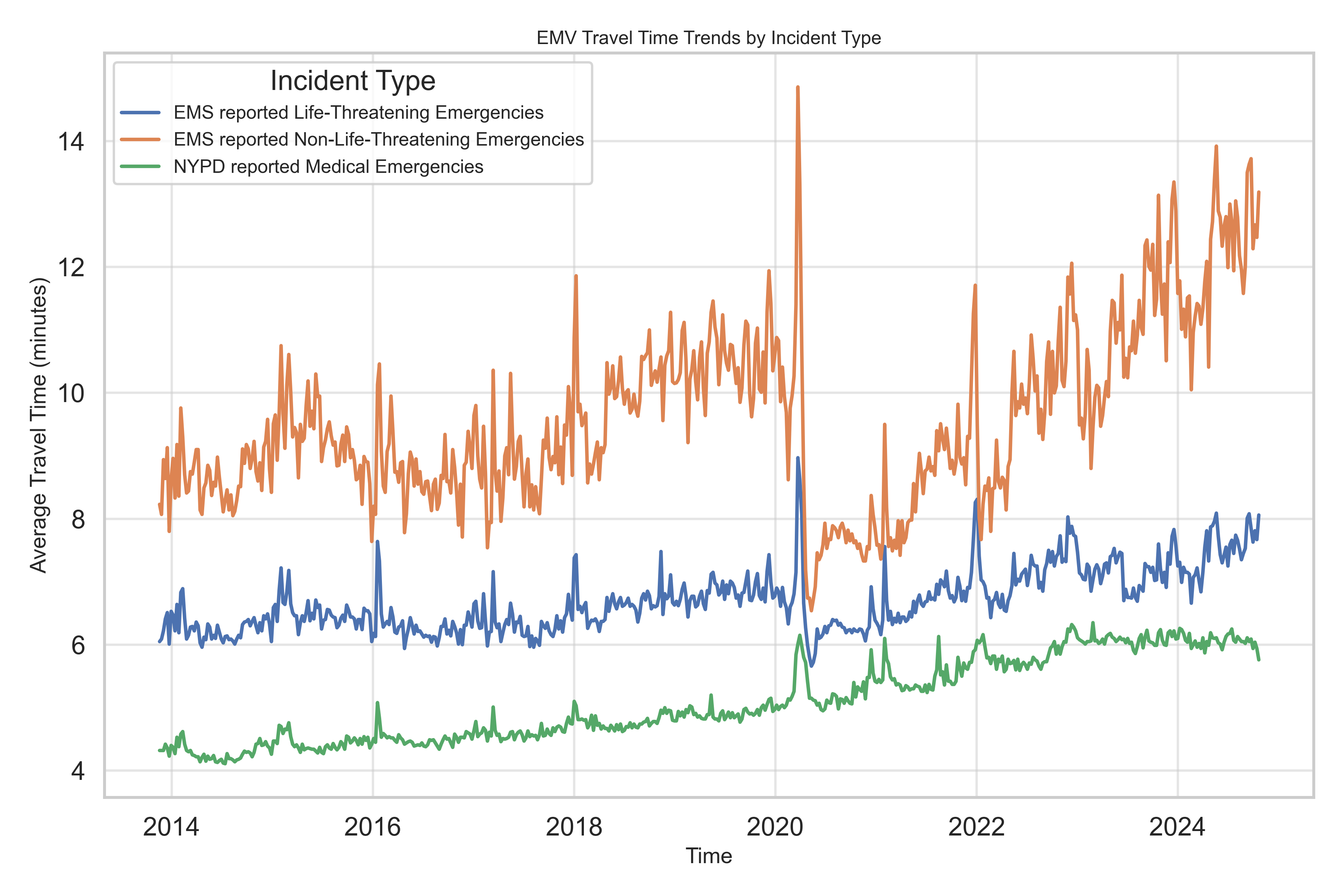}
    \caption{EMV travel time trend based on 911 NYC End-to-End.}
    \label{fig:emv_travel_time_trend}
\end{figure}
For reference, Fig.~\ref{fig:emv_travel_time_trend} illustrates the trend of EMV travel time over the past decade, revealing a continuous increase in EMV travel time for NYPD-reported medical emergencies throughout the study period. Notably, EMV travel time for both life-threatening and non-life-threatening emergencies experienced an abrupt decline during the outbreak of Covid-19 in the city due to shelter-in-place, followed by a gradual rise in the post-Covid-19 period.
\begin{table}
    \centering
    \renewcommand{\arraystretch}{1.2}
    \setlength{\tabcolsep}{6pt} 
    \begin{tabular}{lcc}
        \toprule
        \textbf{} & \textbf{$T_{actual}$ (minutes)} & \textbf{$T_{actual}/T_{simulated}$}\\
        \midrule
        25\% & 6.39 & 2.34\\
        50\% & 7.62 & 1.92\\
        75\% & 9.40 & 1.89\\
        97.5\% & 12.34 & 1.69\\
        100\% & 14.86 & 1.77\\
        Average & 7.96 & 1.80\\
        \bottomrule
    \end{tabular}
    \caption{Statistics for actual travel time ($T_{actual}$) based on 911 End-to-End reports. The ratio $T_{actual}/T_{simulated}$ compares actual travel time to simulated travel time for overall medical service when $\alpha = 15\,\text{s}\cdot\text{m}^2$.}
    \label{tab:real_statistics_travel_time}
\end{table}

\hs{
If we calculate the statistics of the actual EMV travel time $T_{actual}$ for all medical incidents and compare it with the simulated travel time $T_{simulated}$ when $\alpha = 15\,\text{s}\cdot\text{m}^2$, as shown in Table~\ref{tab:real_statistics_travel_time}, we observe that the simulated travel times in Table~\ref{tab:travel_time_summary} tend to underestimate the actual travel times. This observation aligns with the findings of \cite{chung2024access}, as both studies assume that EMVs always travel at the speed limit between intersections. Furthermore, we note that for shorter trips, the ratio $T_{actual}/T_{simulated}$ tends to be larger compared to longer trips. This can be attributed to intersection delay factors, which play a proportionally greater role in shorter trips, whereas the simulation-to-reality gap narrows for longer trips due to the influence of extended road segments.

While it may be tempting to scale the intersection delay factor $\alpha$ by the average underestimation ratio ($1.8\times 15 = 27\,\text{s}\cdot\text{m}^2$), this approach oversimplifies the underlying dynamics of EMV travel times. Firstly, attributing the majority of travel time to intersection delay overlooks the fact that urban travel times are influenced by numerous interacting factors, such as traffic congestion, vehicle acceleration and deceleration, driver behavior, and road geometry~\cite{zhang2007intersection}. These factors are not uniformly distributed across all trips and vary based on location, time of day, and road conditions. Assigning a higher $\alpha$ uniformly would disproportionately amplify the role of intersections and fail to capture these complexities. For instance, segments with fewer intersections but heavy congestion would still exhibit significant delays, which a simple scaling approach cannot account for.

Secondly, the model assumes that intersection delay is additive rather than multiplicative, meaning the travel time for a road segment is modeled as the sum of its free-flow travel time and an intersection delay component. Scaling $\alpha$ directly would imply a shift in this underlying assumption, potentially leading to inconsistent results. Increasing $\alpha$ would exaggerate delays at intersections without considering the diminishing marginal impact of additional intersections, especially in areas where traffic congestion or coordinated signals mitigate delay. This would likely result in unrealistic travel time predictions in regions with high intersection densities but well-managed traffic flow.

Finally, experiments conducted on the $\text{Synthetic Grid}_{5\times5}$ map in Section~\ref{subsec:improvement} demonstrate that $\alpha = 15\,\text{s}\cdot\text{m}^2$ aligns closely with observed delays in scenarios lacking pre-emption. This consistency underscores that $\alpha = 15\,\text{s}\cdot\text{m}^2$ provides a reasonable baseline for evaluating EMS accessibility without overestimating intersection effects. While the model underestimates absolute travel times due to idealized assumptions, $\alpha = 15\,\text{s}\cdot\text{m}^2$ strikes a balance, effectively capturing relative differences in EMS accessibility and identifying vulnerable regions.
}

\subsection{Vulnerable regions}\label{subsec:vulnerable_regions}
\begin{figure}[H]
    \centering
    \includegraphics[width=\textwidth]{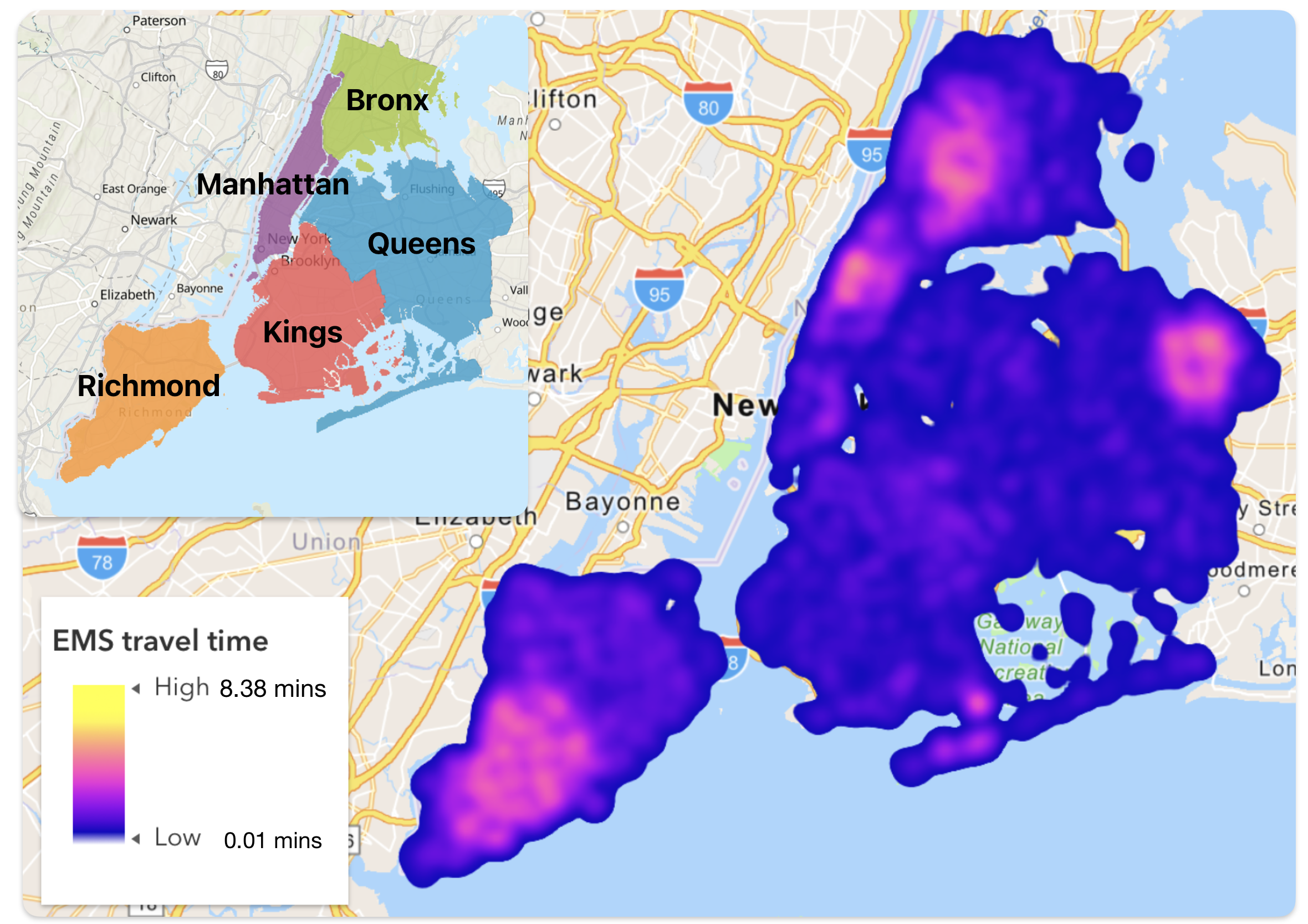} 
    \caption{EMS travel time heatmap, with heated regions highlight\hs{ing} the vulnerable regions.}
    \label{fig:vulnerable_regions}
\end{figure}

Based on the simulated travel time in Subsec.~\ref{subsec:travel_time} for the overall EMS accessibility with $\alpha = 15\text{s}\cdot\text{m}^2$, we here highlight the EMS accessibility vulnerable regions in NYC, see Fig.~\ref{fig:vulnerable_regions}.

We identified four regions within New York City that are particularly vulnerable in terms of EMS accessibility, all exhibiting average EMS travel times exceeding the benchmark of $\tau = 4 \, \text{minutes}$. First, the southern and western parts of Staten Island are characterized by sparse population distribution and a suburban layout with a limited number of medical facilities. This geographic configuration leads to prolonged response times and reduced accessibility for residents. Second, the outer ring of Queens faces a dual challenge of low medical facility density and relatively high intersection density. These geographically expansive areas experience significant gaps in EMS coverage, particularly within densely populated residential neighborhoods. Third, the Upper West Side of Manhattan, a densely populated residential neighborhood, is marked by a dense network of intersections and frequent traffic congestion. These conditions pose substantial barriers to efficient EMS response, further highlighting the region's accessibility challenges. Fourth, large areas of the Bronx, which lack sufficient medical facilities, also suffer from inadequate service coverage. Additionally, minor vulnerable regions with average EMS travel times around $\tau = 4 \, \text{minutes}$ are recognized, such as the Rockaway Peninsula. This area, being geographically distant from the city center, faces notable challenges in accessing general medical services.
\section{Discussion}
\label{sec:discussion}
In this section, we first conduct a sensitivity study on the value of $\tau$ in Subsec.~\ref{subsec:tau}. We examine accessibility' with respect to demographic factors in Subsec.~\ref{subsec:demographic_analysis}, and then further investigate \textit{EMVLight}'s potential to facilitate these EMS accessibility vulnerable area in Subsec.~\ref{subsec:improvement}.

\subsection{Feasibility of \texorpdfstring{$\tau$}{tau}}\label{subsec:tau}
\begin{figure}[ht]
    \centering
    \includegraphics[width=0.75\textwidth]{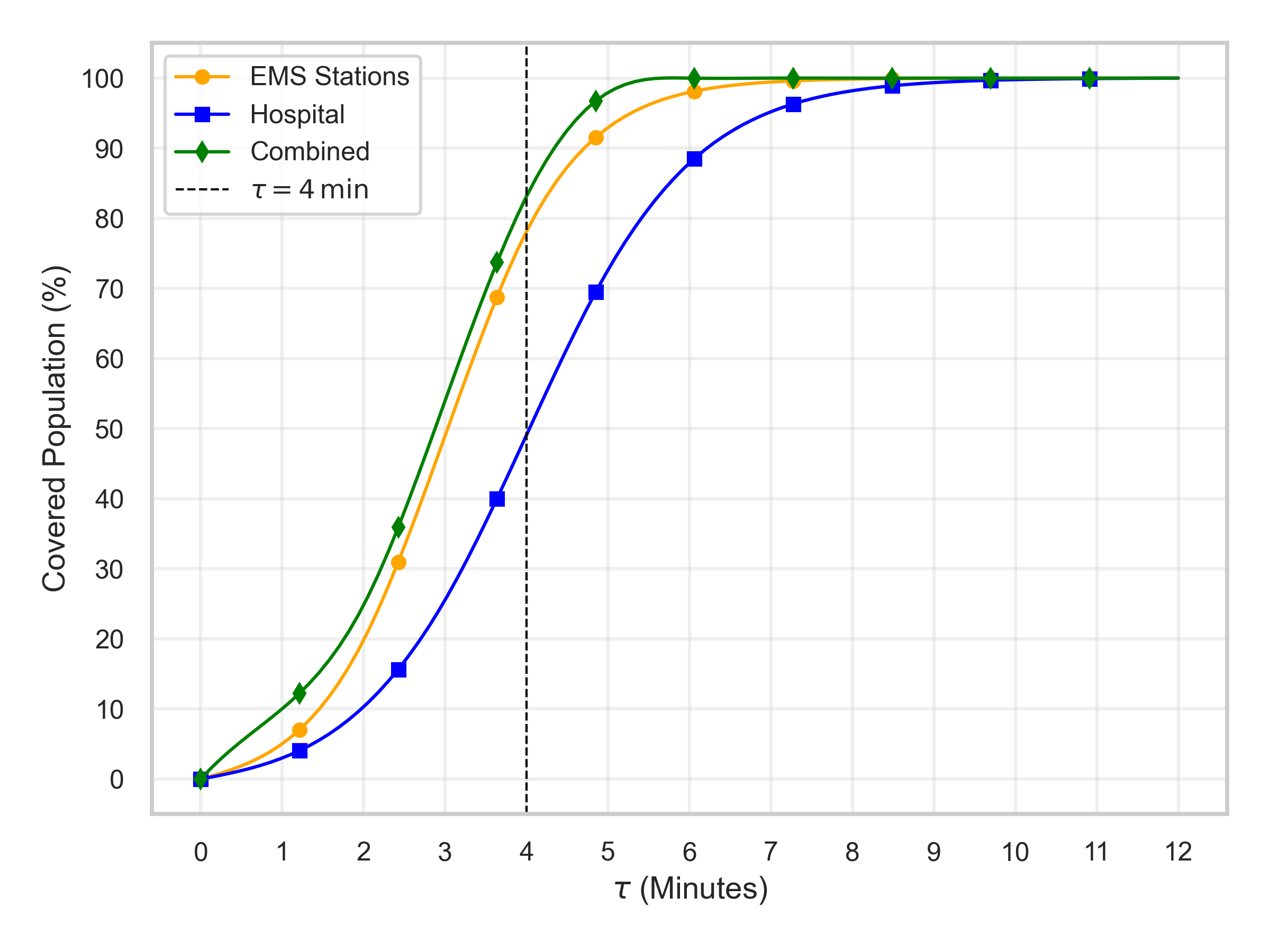} 
    \caption{Population covered by EMS accessibility as a function of $\tau$.}
    \label{fig:covered_func}
\end{figure}

Even though $\tau = 4 \, \text{minutes}$ is suggested by NFPA 1710 standards~\cite{NFPA1710} and widely adopted, its feasibility in NYC remains uncertain. Using the population assignment framework proposed in Subsubsec.~\ref{subsubsec:pop_assignment}, we evaluate the percentage of the population covered by EMS accessibility as a function of benchmark travel time $\tau$. Fig.~\ref{fig:covered_func} illustrates the accessibility of emergency services (EMS stations, hospitals, and combined) as a function of $\tau$. The results indicate a rapid increase in population coverage for smaller $\tau$, followed by a plateau as $\tau$ increases. At $\tau = 4 \, \text{minutes}$, EMS stations cover approximately 80\% of the population, aligning with NFPA 1710 standards, while hospital coverage remains lower, at around 50\%.

This disparity underscores the critical role of EMS stations in ensuring rapid response and highlights the need for infrastructure improvements to enhance hospital accessibility. Beyond $\tau = 4 \, \text{minutes}$, additional gains in coverage diminish, making $\tau = 4 \, \text{minutes}$ a practical benchmark for EMS station accessibility. When EMS stations and hospitals are considered jointly, 83\% of the population can be reached within 4 minutes. However, hospitals, with greater capacity to handle critical emergencies, remain indispensable for specific medical situations. \hs{The current allocation of hospitals suggests that a benchmark travel time of $\tau = 5.5 \, \text{minutes}$ is more realistic for hospital accessibility in NYC, which exceeds the NFPA 1710 requirement of 4 minutes. This finding highlights the urgent need for additional hospital locations to ensure equitable and timely access to critical medical care.}

\subsection{Demographic analysis}\label{subsec:demographic_analysis}
To explore correlations and extend vulnerability analysis to socio-economic dimensions, we compare the identified vulnerable regions with demographic distributions, such as age and median income, derived from the U.S. Census~\cite{bureau2023census}. Figures are available in Appendix~\ref{appendix}.
\subsubsection{Age}
The demographic distribution of NYC reveals significant variations in age structure across census tracts. Fig.~\ref{fig:65+_distribution} highlights that the 65+ population is concentrated in Staten Island and southern Brooklyn, where several census tracts have over 1,700 elderly residents. In contrast, the Bronx and parts of northern Manhattan have notably lower concentrations, with many tracts reporting fewer than 338 individuals aged 65+. Fig.~\ref{fig:median_age_distribution}, depicting median age, complements this observation, showing that Staten Island and southern Brooklyn exhibit the highest median ages (above 50 years), indicative of older populations. Conversely, the Bronx and northern Brooklyn demonstrate significantly younger median ages, often below 30 years. This spatial dichotomy underscores that demographic aging in Staten Island and southern Brooklyn renders these areas more medically vulnerable, as they are likely to demand greater age-specific medical services.

Overlaying the identified EMS-vulnerable regions with demographic data, particularly the distribution of the population aged 65 and over, reveals several critical insights. The Upper West Side of Manhattan stands out with a notably high concentration of elderly residents, a demographic that is especially reliant on timely emergency medical services. This significant proportion of older adults, combined with the area's existing EMS accessibility challenges, amplifies the region’s vulnerability. Similarly, the peripheral areas of Queens exhibit comparable risks, where limited medical infrastructure coincides with a growing elderly population. Staten Island further highlights this intersection of demographic and geographic challenges, particularly in its southern and western census tracts, which feature a high ratio of residents aged 65 and older. As Staten Island is broadly considered an EMS accessibility desert, the substantial presence of seniors in these areas exacerbates the already critical gaps in emergency medical response. These findings underscore the urgent need for targeted interventions to improve EMS accessibility in regions with a high concentration of vulnerable elderly populations.

\subsubsection{Median Income}
We present maps highlighting pronounced socioeconomic disparities across NYC, with Fig.~\ref{fig:median_income_distribution} depicting the median income per census tract and Fig.~\ref{fig:poverty_percentage_distribution} illustrating the poverty population ratio per census tract. High-income areas, primarily located in Manhattan (e.g., Upper East Side, Midtown), exhibit median incomes exceeding \$99,680, which align with low poverty ratios (below 10.8\%). Conversely, the Bronx stands out as the most economically vulnerable borough, characterized by median incomes below \$34,450 and poverty rates exceeding 34.1\%, followed by economically distressed neighborhoods in Brooklyn, such as East New York and Brownsville. This inverse correlation between income and poverty highlights the potential larger medical needs in these underserved areas, particularly in the Bronx and select Brooklyn neighborhoods.

Overlaying the identified EMS-vulnerable regions with median income distribution data reveals critical socioeconomic disparities impacting EMS accessibility. Notably, the peripheral areas of Queens exhibit higher poverty rates compared to surrounding neighborhoods, intensifying the challenges posed by limited medical infrastructure and relatively high intersection densities. In Staten Island, particularly the southern and western regions, residents experience relatively low income levels, compounding the area's designation as an EMS desert and further hindering timely emergency responses. Conversely, while the Bronx registers the lowest median household income among New York City's boroughs, the majority of its areas are not classified as EMS-vulnerable. 

\subsection{Improvement with EMVLight}\label{subsec:improvement}
\textit{EMVLight}~\cite{su2023emvlight} is a state-of-the-art large-scale multi-agent reinforcement learning (MARL) framework designed to optimize real-time TSC for facilitating the efficient passage of EMVs in congested road networks. By dynamically assigning the roles of pre-emption agents to signalized intersections as EMVs traverse the network, \textit{EMVLight} enables real-time shortest path navigation and coordinated signal control, ensuring seamless EMV mobility through highly congested environments.

Experimental evaluations on synthetic maps (e.g., Synthetic Grid\(_{5 \times 5}\)) 
and real-world maps (e.g., Manhattan, \(16 \times 3\); Hangzhou, \(4 \times 4\)) 
demonstrated that \textit{EMVLight} can reduce EMV travel time (\(T_{\text{EMV}}\)) 
by 60\% to 80\%, depending on whether the network is emergency-capacitated, 
compared to benchmarks without pre-emption. These results highlight the 
framework’s ability to substantially enhance EMV traversal efficiency across 
varying network configurations and densities.

To assess \textit{EMVLight}'s ability to reduce $T_{\text{EMV}}$ within the perspective of the EMS accessibility model proposed in Subsec.~\ref{subsec:ems_model}, we can study the amount of time the EMV takes passing through intersections against the benchmark to estimate the potential intersection delay reduction. 

\begin{figure}[H]
    \centering
    \includegraphics[width=\textwidth]{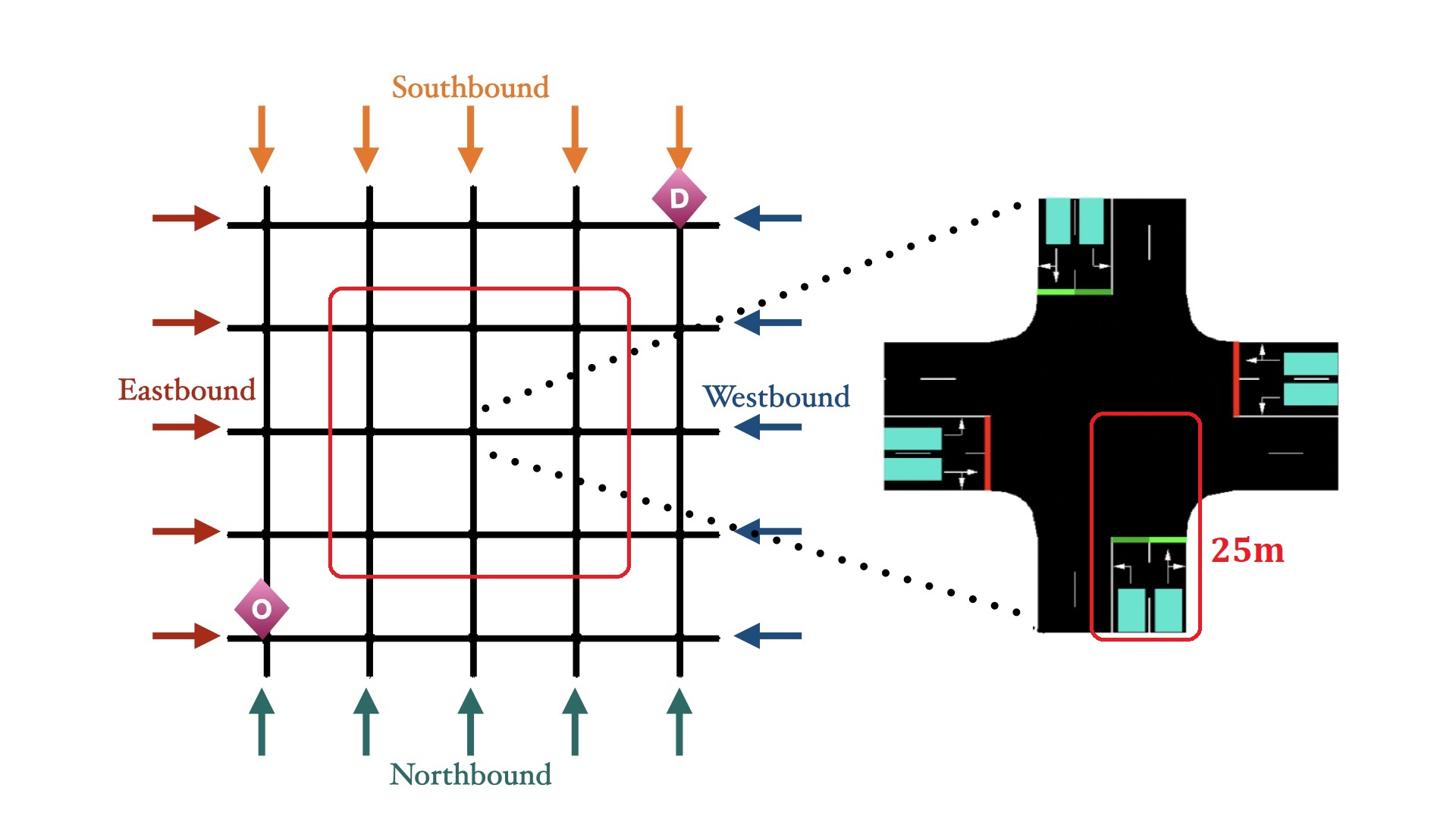} 
    \caption{$\text{Synthetic Grid}_{5\times5}$ and the inner 9 intersections are observed for intersection delay.}
    \label{fig:synthetic_grid}
\end{figure}

Let us revisit the $\text{Synthetic Grid}_{5 \times 5}$ map, shown in Fig.~\ref{fig:synthetic_grid}, where intersections are interconnected through bi-directional links, each consisting of two lanes. For this experiment, we assume that all links have zero emergency capacity. The traffic configuration selected for this study involves northbound and southbound flows transitioning to eastbound and westbound directions, with a non-peak flow of 200 vehicles per lane per hour and a peak flow of 240 vehicles per lane per hour. The origin (O) and destination (D) of the EMV are explicitly labeled on the map to highlight the designated travel route. The traffic flow for this synthetic map spans a duration of 1200 seconds, during which varying congestion levels are simulated. To ensure the network is sufficiently congested, the EMV is dispatched at \( t = 600 \, \text{s} \), a point when traffic conditions have reached compactness.

We record the time taken by the EMV as it traverses any of the inner nine intersections, with the measurement boundaries defined as 25 meters from the center of each intersection in the incoming direction to 25 meters in the outgoing direction. This precise delineation ensures consistency in capturing the traversal times across intersections. The benchmarks for comparison are 1.\textbf{Walabi}~\cite{bieker2019modelling}, which is a rule-based control scheme aiming Green Wave for EMVs in SUMO environment, and 2. no pre-emption for EMVs. 

\begin{figure}[H]
    \centering
    \includegraphics[width=0.9\textwidth]{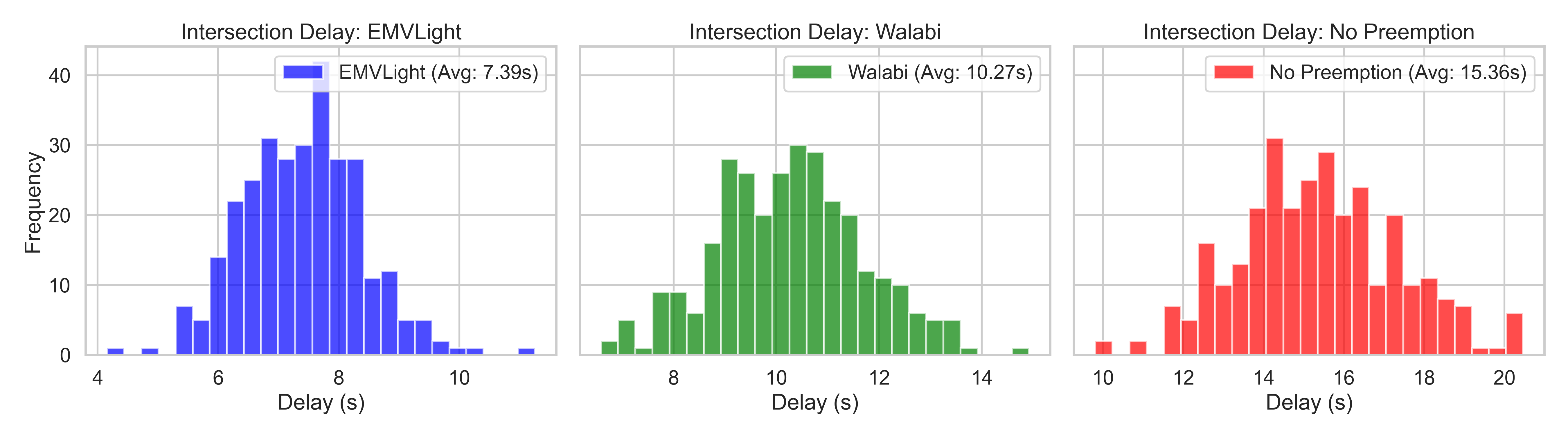} 
    \caption{Amount of time to pass intersections for EMVs.}
    \label{fig:intersection_delay}
\end{figure}

After running the simulations 120 times for each scenario, we aggregate and present the time required for the EMV to traverse the observed intersections in Fig.~\ref{fig:intersection_delay}. The results clearly demonstrate that \textit{EMVLight} reduces intersection delays by approximately 30\% compared to the Walabi and achieves a time savings of over 50\% relative to no pre-emption scenarios. 

\begin{figure}[H]
    \centering
    \includegraphics[width=.75\textwidth]{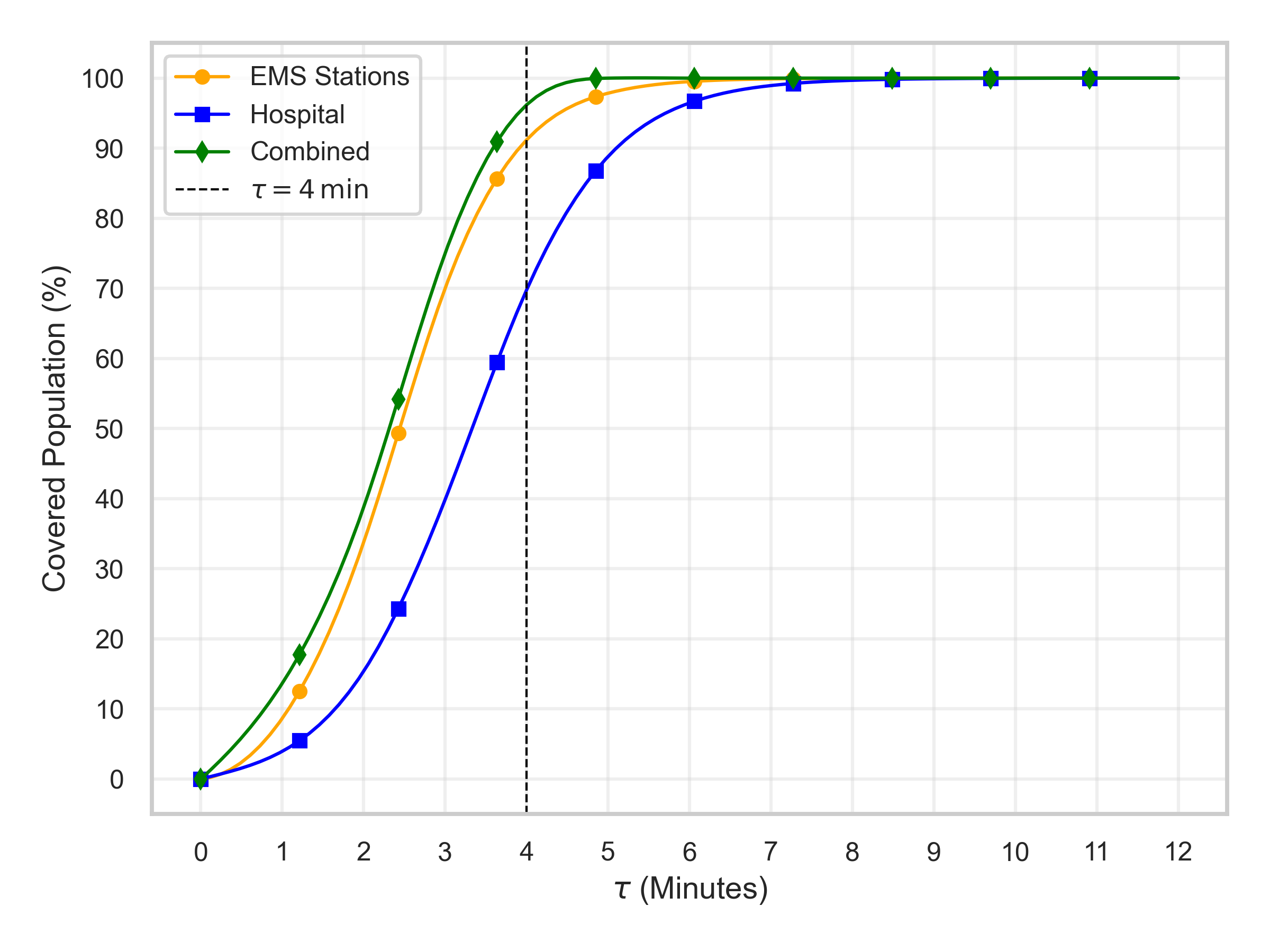} 
    \caption{Population coverage function against $\tau$ with $\alpha = 7.5 \text{s}\cdot\text{m}^2 $}
    \label{fig:emvlight_pop_covered}
\end{figure}

The results indicate that incorporating \textit{EMVLight} across the entire city effectively reduces the intersection delay factor, \(\alpha\), by half. By adopting \(\alpha = 7.5 \, \text{s} \cdot \text{m}^2\), the updated population coverage function, shown in Fig.~\ref{fig:emvlight_pop_covered}, reveals significant improvements in EMS accessibility. Approximately 70\% of the population gains access to hospitals, and 95\% of NYC residents can be served within \(\tau = 4 \, \text{minutes}\). These results highlight the transformative potential of \textit{EMVLight} in enhancing emergency service coverage across the city.

\section{Conclusion}\label{sec:conclusion}
This study presents an intersection-aware EMS accessibility model to comprehensively evaluate medical emergency service coverage in NYC. By incorporating road network characteristics, intersection delays, and demographic data, the model identifies critical EMS-vulnerable regions, including Staten Island, the peripheral regions of Queens, and parts of Manhattan. These findings emphasize the challenges faced by underserved areas, particularly those with limited medical infrastructure, high intersection densities, or significant demographic vulnerabilities. The analysis further demonstrates the capability of \textit{EMVLight} to significantly improve EMS performance by halving intersection delays, increasing hospital accessibility to 70\%, and ensuring 95\% of NYC residents are served within the benchmark travel time of \(\tau = 4 \, \text{minutes}\).

Despite its contributions, the study is subject to certain limitations. The absence of real-time traffic condition data in the travel time assignment likely underestimates congestion-induced delays, and the assumption of free-flow travel between intersections simplifies the complexity of urban traffic dynamics. These constraints highlight the need for future work to incorporate dynamic traffic data and congestion modeling to refine the accessibility framework. Additionally, the static nature of the EMS station locations in the model does not account for adaptive dispatch strategies, which could further enhance emergency response capabilities in underserved regions.

Nonetheless, this study provides a robust foundation for understanding EMS accessibility and exploring the role of advanced TSC systems like \textit{EMVLight} in improving emergency vehicle navigation. The intersection-aware framework and the integration of demographic and network data offer valuable insights into the spatial and temporal disparities in EMS access. Future studies should focus on validating the model using real-world traffic data, exploring the scalability of TSC systems across larger urban networks, and assessing the cost-effectiveness of implementing such systems. By addressing these aspects, the findings of this study can inform policy decisions and urban planning strategies aimed at reducing response times and ensuring equitable access to emergency medical services in cities like NYC and beyond.

\hfill
\FloatBarrier
\printbibliography
\FloatBarrier

\appendix
\section{Appendix}\label{appendix}
In this section, we list the detailed population distribution maps from 2023 US Census~\cite{bureau2023census}. See below.

\begin{figure}[ht]
    \centering
    \includegraphics[width=\textwidth]{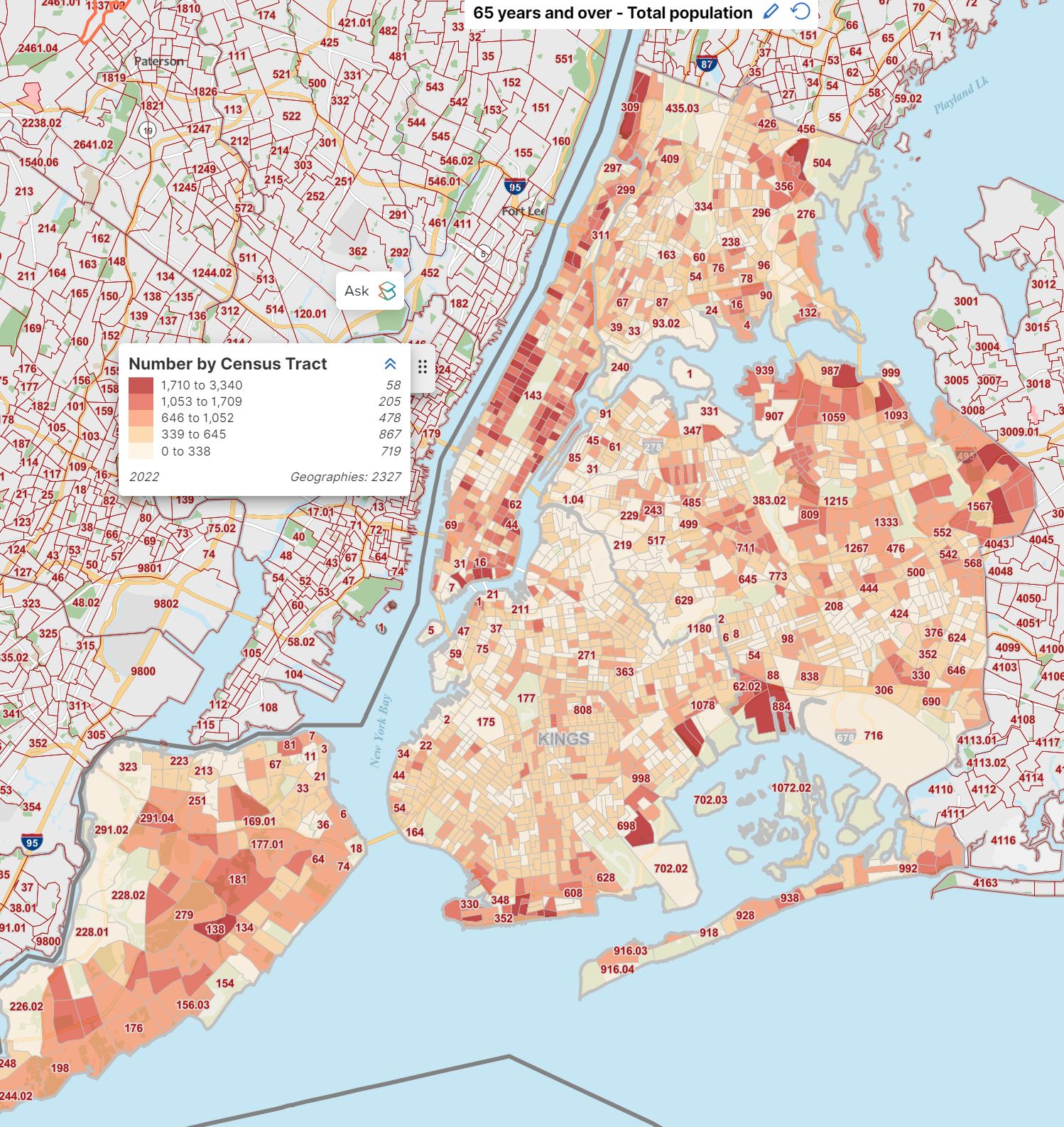} 
    \caption{Population of 65+ years old distribution of NYC census tracts}
    \label{fig:65+_distribution}
\end{figure}
\begin{figure}[ht]
    \centering
    \includegraphics[width=\textwidth]{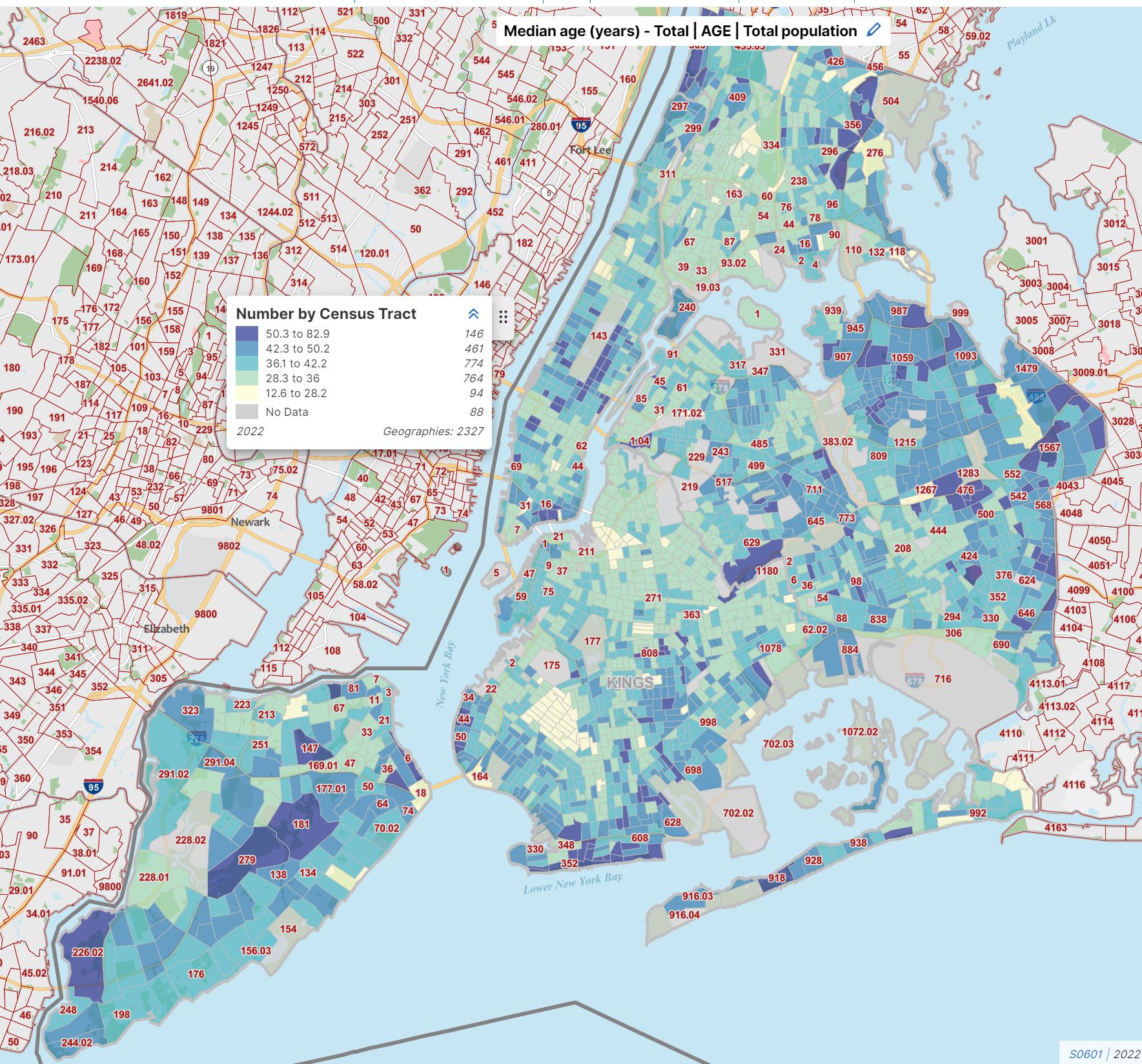} 
    \caption{Median age distribution of NYC census tracts.}
    \label{fig:median_age_distribution}
\end{figure}
\begin{figure}[ht]
    \centering
    \includegraphics[width=\textwidth]{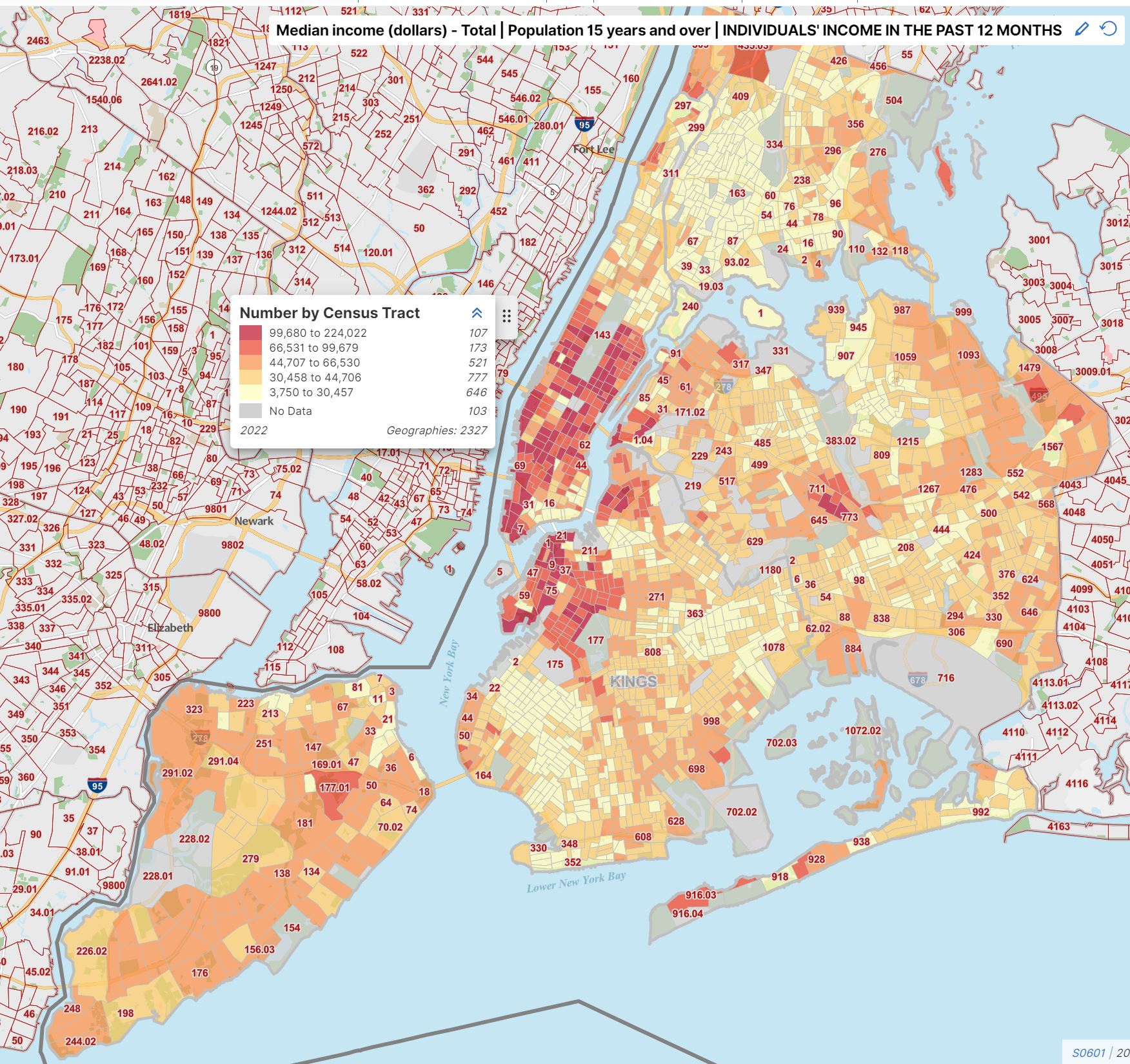} 
    \caption{Median income distribution of NYC census tracts}
    \label{fig:median_income_distribution}
\end{figure}
\begin{figure}[ht]
    \centering
    \includegraphics[width=\textwidth]{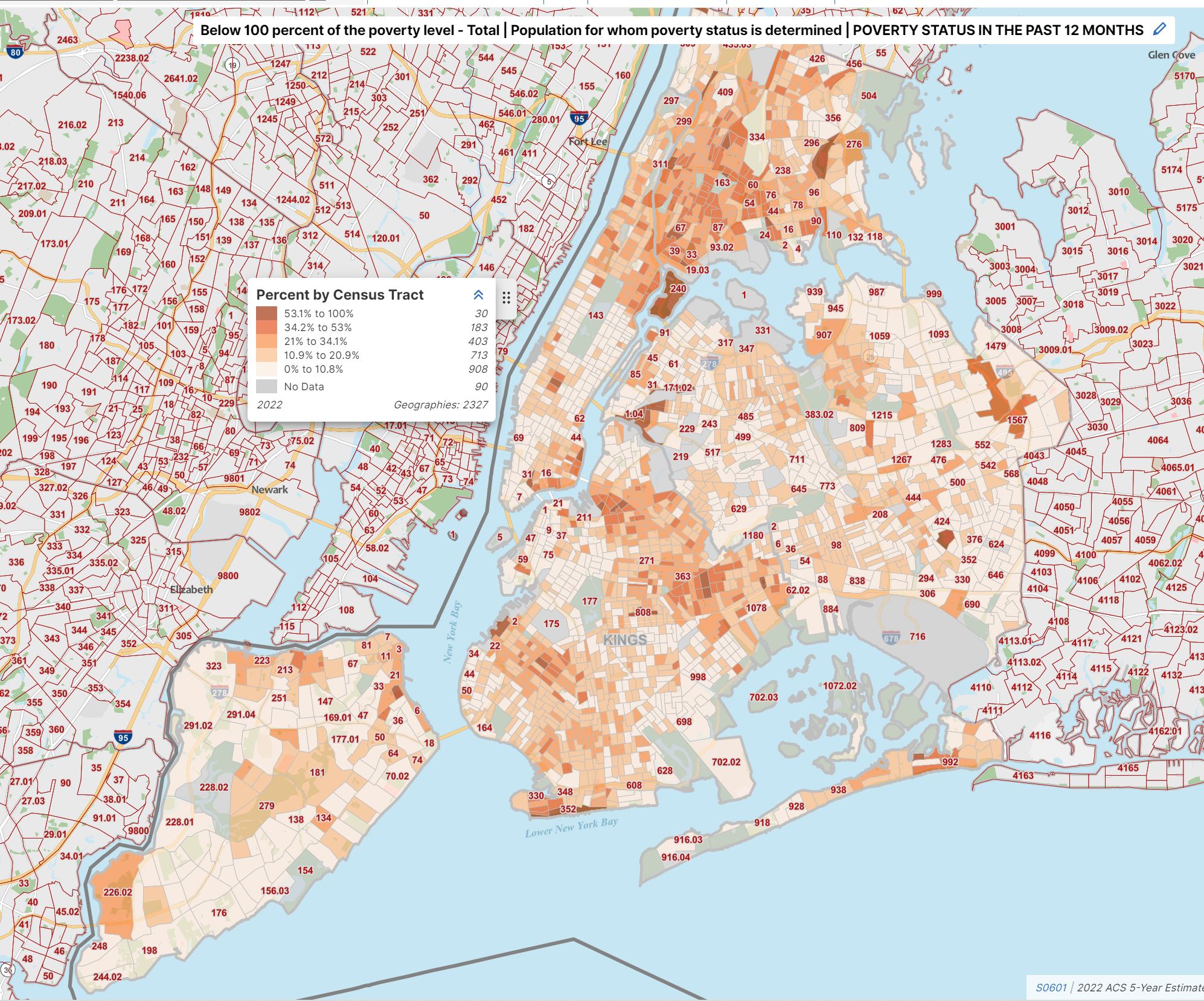} 
    \caption{Poverty percentage distribution of NYC census tracts}
\label{fig:poverty_percentage_distribution}
\end{figure}
\end{document}